\documentclass[11pt]{article}
\usepackage[left=3cm, right=3cm,top=3cm, bottom=3cm]{geometry}

\usepackage{moreverb,url}

\usepackage[colorlinks,bookmarksopen,bookmarksnumbered,citecolor=red,urlcolor=red]{hyperref}

\usepackage{bm, amsmath, bbold, amssymb, array, lipsum}
\usepackage{mathtools}
\usepackage{setspace}
\usepackage{placeins}
\usepackage{multirow}
\usepackage{ctable}
\usepackage{xcolor, soul}
\usepackage{natbib}
\usepackage{graphicx}
\usepackage{enumitem}
\graphicspath{{./pics/}}
\usepackage{appendix}
\usepackage{soul}
\usepackage{textcomp}

\usepackage{authblk}

\usepackage[T1]{fontenc}
\usepackage[utf8]{inputenc}

\usepackage{amsmath}

\newcommand{\keywords}[1]{ {\bf\it Keywords:} #1}

\newcommand{\bzero}{ { \mathbf{0} }}

\newcommand{\bsb} { {\boldsymbol{b}} }

\newcommand{\bI}{ { \bf I }}

\newcommand{\bsr}{ { \boldsymbol{r} }}

\newcommand{\bR}{ { \bf R }}

\newcommand{\bw}{ { \bf w }}
\newcommand{\bW}{ { \bf W }}

\newcommand{\bX}{ { \bf X }}

\newcommand{\bY}{ { \bf Y }}

\newcommand{\bZ}{ { \bf Z }}

\newcommand{\bsW}{ { \boldsymbol{W} }}
\newcommand{\bsw}{ { \boldsymbol{w} }}

\newcommand{\bsx}{ { \boldsymbol{x} }}
\newcommand{\bsY}{ { \boldsymbol{Y} }}
\newcommand{\bsy}{ { \boldsymbol{y} }}

\newcommand{\beps}{ { \boldsymbol{\epsilon} }}
\newcommand{\btheta}{ { \boldsymbol{\theta} }}

\newcommand{\bmu}{ { \boldsymbol{\mu} }}

\newcommand{\bPsi}{ { \boldsymbol{\Psi} }}
\newcommand{\bbeta}{ { \boldsymbol{\beta} }}

\newcommand{\bSigma}{ { \boldsymbol{\Sigma} }}
\newcommand{\bgamma}{ { \boldsymbol{\gamma} }}

\newcommand{\bOmega}{ { \boldsymbol{\Omega} }}
\newcommand{\bLambda}{ { \boldsymbol{\Lambda} }}
\newcommand{\ind}{ \mathbb{I} }
\newcommand{\Exp}{ \mathrm{E} }

\newcommand{\pr}{ {\text{pr}} }
\newcommand{\MVN}{ {\text{MVN}} }
\newcommand{\MVt}{ {\text{MVt}} }
\newcommand{\logit}{ {\text{logit}} }
\newcommand{\IW}{ {\text{IW}} }
\newcommand{\Var}{ {\text{Var}} }

\newcommand{\homhov}{\textit{HOM-HOV} }
\newcommand{\hemhov}{\textit{HEM-HOV} }
\newcommand{\homhovo}{\textit{HOM-HOV-O}}
\newcommand{\hemhovo}{\textit{HEM-HOV-O}}
\newcommand{\hemhevo}{\textit{HEM-HEV-O}}
\newcommand{\tR}{t\text{-R}}
\newcommand{\tE}{t\text{-E}}
\newcommand{\tRE}{t\text{-RE}}

\newcommand{\WAIC}{ \text{WAIC} }
\newcommand{\RIL}{ \text{RIL}}

\bibpunct{(}{)}{;}{a}{,}{,}

\definecolor{dark-gray}{gray}{0.26}
\definecolor{light-gray}{gray}{1}

\definecolor{darkgreen}{HTML}{009F00}

\newcommand\BibTeX{{\rmfamily B\kern-.05em \textsc{i\kern-.025em b}\kern-.08em
T\kern-.1667em\lower.7ex\hbox{E}\kern-.125emX}}

\usepackage{xcolor}
\definecolor{custom_bg}{RGB}{60,68,76}

\begin{document}

\title{Covariate Informed Identification of Heterogeneity and Outliers in Longitudinal Data}

\author[1]{Anish Mukherjee}
\author[1,*]{Jeremy T. Gaskins}
\author[2]{Jeffrey M. Albert}
\author[3]{Tanujit Dey}
\author[4]{Sounak Chakraborty}
\affil[1]{Department of Bioinformatics and Biostatistics, University of Louisville, Louisville, Kentucky, USA}
\affil[2]{Department of Population and Quantitative Health Sciences, Case Western Reserve University, Cleveland, Ohio, USA}
\affil[3]{Center for Surgery and Public Health, Mass General Brigham, Harvard Medical School, Boston, Massachusetts, USA}
\affil[4]{Department of Statistics and Data Science, University of Missouri, Columbia, Missouri, USA}
\affil[*]{Corresponding author: Jeremy Gaskins. Email: \texttt{jeremy.gaskins@louisville.edu}}

\date{}

\maketitle

\begin{abstract}
    We often observe heterogeneity in longitudinal data, where the mean and variance for certain profiles meaningfully differ from the rest. 
    Some profiles may also exhibit outliers at a limited number of measurements. 
    Using a standard mixed effects model, which assumes homogeneity, can lead to overestimating the residual variance and inefficient estimation. 
    In this work, we identify and account for three sources of heterogeneity in longitudinal data: incompatible mean trajectories, increased residual variance, and outliers at individual measurements. 
    Our Bayesian mixture model incorporates binary indicators of heterogeneity for each of these features, modeled through logistic regression using covariates. 
    We perform statistical inference using Markov chain Monte Carlo and implement model selection to evaluate the inclusion of various heterogeneous components. 
    Simulations demonstrate that our model can accurately identify heterogeneity and produce efficient estimates of the fixed effects parameters. 
    We further validate our approach using the DHEAS hormone data from the SWAN study.
\end{abstract}



\section{Introduction} 
\label{sec:intro}

Mixed-effects models are the standard for analysis of longitudinal data \citep{lairdRandomEffectsModelsLongitudinal1982, chiModelsLongitudinalData1989}, and are typically structured as $\bsY_i = \bX_i \bbeta + \bZ_i \bsr_i + \beps_i$. 
In this framework, population-level fixed effects $\bbeta$ and individual random effects $\bsr_i \sim \MVN(\bzero, \bLambda)$ determine the profile mean, while errors $\beps_i \sim \MVN(\bzero, \bOmega)$ capture residual noise. 
However, this simple specification often fails to account for inherent heterogeneity. 
Our goal in this project is to develop an estimation methodology that, in addition to being robust, can also identify observations that deviate from the general population structure.
Here, we use {\it heterogeneity} to refer to a structure where there is a single homogeneous group alongside a collection of non-members (i.e., outliers) that each have their own behavior separate from the homogeneous group and from each other. 
This is different from other uses of heterogeneity to refer to a setting with multiple clusters of observations where members of each cluster has the same general behavior \citep{quintanaBayesianNonparametricLongitudinal2016, yuMixtureRegressionLongitudinal2022}.

To construct our methodology, we define three distinct types of heterogeneity that may be of interest. 
First, a profile may be ``mean heterogeneous'' at the subject-level, when it deviates significantly from the population average $\bX_i\bbeta$ due to an extreme random effect $\bsr_i$ inconsistent with the assumed $\MVN(\bzero, \bLambda)$ distribution. 
For example, a patient in a depression trial might show a consistently higher baseline 
or a steeper decline in symptoms 
than what the fixed effects predict.
Second,  we consider the presence of classical ``observation-level outliers,'' where a single measurement $Y_{ij}$ in profile $\bsY_{i}$ is extreme relative to the random effect $\bsr_i$ due to a large specific error $\epsilon_{ij}$. 
Finally, we address ``variance heterogeneity,'' where many components of $\beps_i$ are inconsistent with $\MVN(\bzero, \bOmega)$. 
Unlike isolated outliers, this represents a profile with residuals that are systematically more variable than predicted by the residual variance for the (homogeneous) population -- such as a patient whose symptoms fluctuate wildly around their individual trend $\bX_i\bbeta + \bZ_i\bsw_i$. 
We seek to develop a model that can accommodate and identify observations subject to all three sources of heterogeneity.

In the literature one option for addressing extreme observations 
is to replace
normal distributions with thick-tailed alternatives. 
For instance, \cite{langeRobustStatisticalModeling1989} and \cite{wakefieldBayesianAnalysisLinear1994} utilized multivariate $t$-distributions for data and random effects, respectively. 
\cite{welsh13ApproachesRobust1997} extended this to model both mean and variance heterogeneity, although they assumed a diagonal $\bLambda$ and shared residual variance. 
While \cite{pinheiroEfficientAlgorithmsRobust2001} later allowed for correlated random effects and individual degrees of freedom, their model forced a uniform level of heterogeneity across mean and variance components and ignored observation-level outliers. 
Notably, a few outliers can inflate residual variance, mimicking variance heterogeneity. 
In the simpler random intercept  setting, \cite{mccullochImprovingPredictionsWhen2023, mccullochFlaggingUnusualClusters2024} proposed estimators that can identify extreme clusters and estimate the corresponding random intercepts without over-shrinking.

An alternative line of research uses Tukey-Huber contamination models for robust estimation of $\bbeta$. 
The Classical Contamination Model \citep[CCM;][]{huberRobustEstimationLocation1964} assumes that most profiles have normal errors while a few are highly noisy, and CCM bounds the influence of outliers on the estimation of fixed effects. 
\cite{richardsonBoundedInfluenceEstimation1997} extended this to mixed models using weight functions to regulate outliers. 
However, these methods typically address only subject-level mean heterogeneity. 
To avoid downweighting an entire profile due to a few extreme points, \cite{alqallafPropagationOutliersMultivariate2009} introduced the Independent Contamination Model (ICM), which treats observations as independent potential outliers. 
\cite{agostinelliCompositeRobustEstimators2016} later proposed composite estimators covering both CCM and ICM, while \cite{kangRobustEstimationLongitudinal2020} developed an efficient Hellinger distance-based estimator for longitudinal data. 
Despite these advances in robust estimation, these frameworks lack a convenient mechanism to identify observations specifically subject to all three sources of heterogeneity.

The contamination class of priors was formally introduced for robust Bayesian analysis by \cite{bergerRobustBayesEmpirical1986}, who studied posterior sensitivity as the prior varies over such a class. 
\cite{bergerRobustBayesianAnalysis1990, bergerOverviewRobustBayesian1994a} further developed global and local sensitivity diagnostics within this framework, while \cite{morenoBayesianRobustnessHierarchical1993} extended the analysis to hierarchical settings, demonstrating that posterior inference can be acutely sensitive to the hyperparameters of the base prior. 
At the data level, \cite{bayarriBayesianMeasuresSurprise2003} formulated outlier detection as a testing problem in which outlying observations arise from a contaminating distribution distinct from that of the majority, without requiring estimation of outlier proportions. 
More recent geometric approaches using Rényi divergence and the Fisher–Rao metric have also been proposed to quantify robustness over contamination classes \citep{kurtekBayesianSensitivityAnalysis2015, al-labadiMeasuringBayesianRobustness2021}. 
Despite this rich literature, no existing framework allows the contamination probability to vary as a function of individual-level covariates, nor accommodates the distinction between observation-level and subject-level outliers that arises naturally in longitudinal data.

In this work we propose an integrated approach to modeling all three sources of heterogeneity following a similar strategy to the contamination models.
To enable identification of each heterogeneity types, we introduce three indicator variables characterizing the profiles with extreme subject-level means, extreme residual variances, as well as observation-level outliers.
We consider individual-specific residual variances, with a mixture of point mass representing the homogeneous group and a dispersed prior, to account for heterogeneous covariance structure and also connect it also to the variance of the random effects to allow for heterogeneity in the mean structure. 
We assume that majority of the data is homogeneous in all aspects, with few cases of at least one heterogeneity type. 
We consider a global residual variance shared across all individuals and
different heterogeneity types are modeled by inflating the corresponding variances with suitable scaling factors. 
Additionally, we associate the indicator variables with covariates to learn the features associated with each type of heterogeneity.

The rest of the article is arranged as follows.
In Section \ref{sec:model} we describe our proposed model for Heterogeniety and Outlier Identification for Longitudinal Data (HOILD).  The computational strategies for posterior estimation and inference are discussed in Section \ref{sec:inf}.
Section \ref{sec:simulation} presents  simulation studies demonstrating the effectiveness of our approach.
Section \ref{sec:real_data} illustrates the effectiveness of our proposed approach using longitudinal hormone data. 
The article concludes with a discussion in Section \ref{sec:discussion} that summarizes the applicability of our approach, potential extensions and future directions.

\section{Heterogeniety and Outlier Identification for Longitudinal Data (HOILD)}
\label{sec:model}

For $i = 1, \ldots, n$ and $j = 1, \ldots, n_i$, let $Y_{ij}$ be the observation for the $i$-th individual at the $j$-th time-point and $\boldsymbol{Y}_i = (Y_{i1}, \ldots, Y_{in_i})'$ represent the full longitudinal outcome vector for $i$-th individual. 
We allow the time-points at which the observations are recorded to vary across individuals, and for individual $i$, $\boldsymbol{t}_i = (t_{i1}, \ldots, t_{in_i})'$ represent  the vector of observed time-points. 
Let $\mathbf{X}_i$ be the corresponding design matrix of dimension $n_i \times p$ consisting of $p$ covariates for modeling the overall mean structure. 
Typically, this will include a function of the $\boldsymbol{t}_i$.
Further, let $\bZ_i$ denote the $n_i \times q$ design matrix corresponding to the $q$-dimensional random effects.
To account for the heterogeneity in mean and covariance, we consider the following model 
\begin{equation} \label{eq:model_y_eps}
\bsY_i = \bX_i \bbeta + \bZ_i \bsb_i + \beps_i, \quad \beps_i \sim \MVN(\bzero,\sigma_i^2 \bOmega_i).
\end{equation} 
Here, $\boldsymbol{\beta}$ represents the regression coefficients for the fixed effects,  $\bsb_i$  the random effects, and $\sigma_i^2$  the individual-specific residual variance.
A standard choice for $\bOmega_i = (r_{i;jk})_{1 \leq j,k \leq n_i}$ would be the identity matrix or a first-order auto-correlation matrix.
Correlation matrices with different temporal dependence specifications have been briefly discussed in \cite{zhangSemiparametricStochasticMixed1998}.

\subsection{Indicators characterizing heterogeneity}  \label{sec:model_heterogeneity}


We control the heterogeneity in the HOILD mixed effect model (\ref{eq:model_y_eps}) through our distributional choices  for the patient-specific random effects $\bsr_i$ and the residual covariance $\sigma_i^2\bOmega_i$.  To that end, we will introduce a sequence of indicator variables for each of our three sources of heterogeneity and construct models for the relevant parameters through mixture models.
For every source of heterogeneity, we will specify a contaminated model \citep{huberRobustEstimationLocation1964, gleasonUnderstandingElongationScale1993}, which is essentially a mixture of homogeneous model and a heterogeneous model with inflated covariance structure to accommodate extreme data.

It is natural to assume that the majority of  profiles $\bsY_i$ will be similar in terms of their random effects characterizing a homogeneous group compared with the heterogeneous rest who have more extreme random effects. 
We identify this mean heterogeneity by introducing an indicator variable $U_i$ for $i$-th individual and model the random effects as 
\begin{equation} \label{eq:model_mean}
\bsb_i|U_i=u_i \sim \MVN \left(\bzero, (\eta_u^2)^{u_i} \sigma_i^2 \bLambda \right),
\end{equation}
where $\bLambda$ represents the global covariance matrix for the homogeneous random effects, scaled by the individual specific residual variance $\sigma_i^2$.
Note that $U_i = 0$ indicates the homogeneous group with the covariance given by $\sigma_i^2 \bLambda$, while $U_i = 1$ indicates the heterogeneous group where the covariance matrix is scaled up by a factor $\eta_u^2$ ($\eta_u > 1$).
The data model marginalized over the random effects is given by
\begin{equation} \label{eq:model_outlier}
    \bsY_i | U_i = u_i \sim \MVN\left(\bX_i \bbeta, \sigma_i^2 \left\{ (\eta_u^2)^{u_i} \bZ_i \bLambda \bZ_i'+ \bOmega_i \right\} \right),
\end{equation}
where the first part of the covariance structure is the contribution from the random effects and the second part is the residual covariance.
When $u_i=1$, the covariance matrix of $\bsY_i$ is inflated (relative to observations with $u_i=0$) through the increase in the leading term, indicating that the entire trajectory can differ substantially from the population average $\bX_i \bbeta$.
Note that we connect the individual-specific residual variances $\sigma_i^2$ to the covariance of the random effects to be able to compute the data likelihood marginalized over $\sigma_i^2$ in closed form, helping posterior computation.
The model specification in (\ref{eq:model_outlier}) as a mixture of a homogeneous model and a heterogeneous model with covariance structure inflated by $\eta_u^2$ resembles the scale contamination model proposed in \cite{gleasonUnderstandingElongationScale1993, mccullochFlaggingUnusualClusters2024}.


Despite accounting for the heterogeneity at the mean level, there may also be some extreme measurements with regard to the corresponding profile residual variance $\sigma_i^2$.
Such observation-level outliers, if not accounted for separately in the model, may lead to overestimation of $\sigma_i^2$ 
and/or poor estimation of $\bbeta$ by operating as influential or leverage points.
To identify such observation-level outliers, we introduce an observation-level latent indicator variable $W_{ij}$ in our modeling approach such that $W_{ij} = 1$ indicates that $Y_{ij}$ is an extreme/outlier observation, and zero otherwise. 
For individual $i$, the residual variance of $Y_{ij}$ is $\sigma_i^2$, the subject-level variance when $W_{ij} = 0$, and is inflated by a factor $\eta_w^2$ ($\eta_w>1$) to be $\eta_w^2 \sigma_i^2$ when $Y_{ij}$ is an outlier ($W_{ij} = 1$). 
Therefore, $\bOmega_i$ in the linear model (\ref{eq:model_outlier}) is given by,
\begin{equation} \label{eq:Omega}
\mathbf{\Omega}_i(\bsW_i, \rho) = (\text{I}-\mathbf{W}_i + \eta_w \mathbf{W}_i) \ \bR_i(\rho) \ (\text{I}-\mathbf{W}_i + \eta_w \mathbf{W}_i),
\end{equation}
where $\bW_i = \text{diag}(\bsW_i)$ is a diagonal matrix and $\bR_i = (r_{i;jk})_{jk}$, parameterized by $\rho$, provides the correlation structure among the residuals for the $i$-th individual.
Henceforth, we will denote $\bOmega_i(\bsW_i, \rho)$ simply by $\bOmega_i$ implicitly assuming the dependence of $\bOmega_i$ on $\bsW_i$ and $\rho$.
As a standard choice, we here assume that $\bR_i$ provides an auto-correlation structure of order one with $r_{i;jk} = \rho^{|t_{ij} - t_{ik}|}$, and $\rho$ is the auto-correlation coefficient, 
although one could  make the alternative choices such as  residual independence conditionally on $\bsr_i$.


Having accounted for subject-level and observation-level outliers, the residual variances for the profiles may still have  heterogeneity at the subject-level, that is,  some individuals may have longitudinal profiles with extreme residual fluctuations compared to the majority of the population.
To that end, we introduce another indicator variable $Z_i$ that characterizes the heterogeneity in $\sigma_i^2$.
The group of $\bsY_i$ with $Z_i = 0$ represent the homogeneous variance group and consist of  profiles having the global residual variance
$\sigma_0^2$. 
Patients who are variance heterogeneous ($Z_i=1$) are believed to have a (subject-specific) residual variance $\sigma_i^2$ that is systematically higher than the global value $\sigma_0^2.$  To that end, we assume these $\sigma_i^2$ are dispersed around $\sigma_1^2$, and 
further assume that $\sigma_1^2$ is centered around $\eta_z^2 \sigma_0^2$ with $\eta_z > 1$.
That is, the heterogeneous group is expected to have  an average variance $\sigma_1^2$ that is larger than the homogeneous group variance $\sigma_0^2$ by a factor of $\eta_z^2$.
The prior for $\sigma_i^2$ is chosen to be a mixture of point mass representing the homogeneous profiles and a dispersed gamma prior modeling the heterogeneous group 
\begin{equation} \label{eq:resid_var_prior}
(\sigma_{i}^2 | Z_i) \sim (1-Z_i) \delta_{\sigma_0^2} +
Z_i \text{Ga} 
\left( \frac{1}{\alpha^2}, \frac{1}{\alpha^2\sigma_1^2} \right).
\end{equation}
Note that we use the gamma parameterization such that when $Z_i=1$, $\sigma_i^2$  has mean $\sigma_1^2$ and  variance $\alpha^2\sigma_1^4$. 
Large $\alpha$ is associated with greater differences in the residual variance across heterogeneous observations.  A smaller $\alpha$ indicates the heterogeneous $\sigma_i^2$s will be more concentrated around $\sigma_1^2$, but as $\sigma_1^2$ is expected to be an $\eta_z$-multiple of $\sigma_0^2$, this clusters the residual variances at a value greater than the global variance.


An advantage of choosing a gamma prior for $\sigma_i^2$ over an inverse-gamma
prior is that in the latter case the shape parameter value must take values in a restricted range for the distribution to have finite moments for $\sigma^2_i$. 
On the other hand,  
the data likelihood under the gamma distribution of $\sigma_i^2$
can be marginalized over the subject-specific variance  resulting in a closed-form distribution for individuals within the heterogeneous group. 
For the data model given in (\ref{eq:model_outlier}) with the mixture prior structure for $\sigma_i^2$ given in (\ref{eq:resid_var_prior}), the data likelihood
marginally over the random effects $\bsb_i$ and the variance $\sigma^2_i$ (conditionally on the heterogeneity indicators) turns out to be a two-component mixture, 
\begin{equation} \label{eq:f0_f1_mixture}
\begin{array}{rcl}
f(\bsy_i|\bbeta, U_i, \bsW_i, Z_i, \sigma_0^2, \sigma_1^2, \rho) & = & (1-Z_i) f_0(\bsy_i|\bbeta, \sigma_0^2, \rho) + Z_i f_1(\bsy_i|\bbeta, \sigma_1^2, \alpha, \rho), \\[2pt]
f_{0}(\bsy_i|\bbeta, U_i, \bsW_i, \sigma_0^2, \rho) & = & \MVN_{n_i}(\bX_i \bbeta, \sigma_0^2 \widetilde{\bOmega}_i) \\[2pt]
f_1(\boldsymbol{y}_i | \boldsymbol{\beta}, U_i, \bsW_i, \sigma_1^2, \rho, \alpha) & = &
\frac{ \left(\alpha^2\sigma_1^2 \right)^{-\zeta/2-n_i/2} |\widetilde{\bOmega}_i|^{-1/2}}{2^{n_i/2+\zeta/2-1}\pi^{n_i/2} \Gamma(1/\alpha^2)} 
S^{\zeta}(\bsy_i) 
K_{\zeta} \left( \frac{\sqrt{2}}{\alpha \sigma_1} S(\bsy_i)  \right),
\end{array}
\end{equation}
where 
$f_0$ and $f_1$ represent the the densities for the homogeneous and heterogeneous variance groups, respectively.  The other terms in   these distributions are given by
$\zeta = 1/\alpha^2 - n_i/2$, $\widetilde{\bOmega}_i = (\eta_u^2)^{u_i}\bZ_i  \bLambda \bZ_i' + \bOmega_i$, $S^2(\bsy_i) = (\boldsymbol{y}_i - \mathbf{X}_i \boldsymbol{\beta})' \widetilde{\bOmega}^{-1} (\boldsymbol{y}_i - \mathbf{X}_i \boldsymbol{\beta})$ representing the correlation adjusted SSE, and $K_\nu(\cdot)$ as the modified Bessel function of the second kind. 
We note that $\bsy_i | z_i=1$ follows 
a location-shifted version of the symmetric-special case of the generalized asymmetric Laplace distribution \citep[GAL;][]{kozubowskiMultivariateGeneralizedLaplace2013}.  That is, conditionally on $Z_i=1$,
$\bY_i = \bX_i\bbeta + \beps_i$ with $\beps_i \sim \text{GAL}(\bSigma=\alpha^2 \sigma_1^2\widetilde{\bOmega}, \bmu=\bzero, s=1/\alpha^2)$.

\subsection{Covariate-informed indicator modeling} \label{sec:model_indicators}

We believe the heterogeneity indicators may be informed by a set of predictor variables, and we wish to model them using predictors.
For instance, in the example from the introduction of a depression trial, there may be measured characteristics such as age, length of condition or baseline symptom level that might influence how likely it is that an individual has an irregular overall trajectory (mean heterogeneity), extreme values at individual measurements (outliers), and/or less predictable patterns (variance heterogeneity).
Learning the associations to relevant characteristics will help in identifying different heterogeneity types.

As the mean and variance indicators $U_i$ and $Z_i$ are at the subject level and do not vary across time, they may be informed only by time-invariant predictors or predictor values at the baseline.
In contrast, the observation-level outlier indicators $W_{ij}$ can be associated with time-varying covariates since there are unique indicators for each measurement occasion $t_{ij}$.
To learn the association of predictors with mean heterogeneity, we fit logistic regression model with $U_i$ as the outcome.
Let $\bX_{u}$ denote the design matrix with the $i$-th row $\bsx_{u;i} = (x_{u;i1}, \ldots, x_{u;id_u})$ consisting of the predictor values for $i$-th individual at baseline.
The logistic model is given by
\begin{equation} \label{eq:logistic_u}
\begin{array}{rcl}
\logit(\pi_{u;i}) & = & \bsx_{u;i}' \bgamma_{u},
\end{array}
\end{equation}
where $\pi_{u;i} = \pr(U_i = 1)$ is the probability that $i$-th profile belongs to the mean heterogeneous group.
Similarly, to identify predictors associated with variance heterogeneity, we fit
\begin{equation} \label{eq:logistic_z}
\begin{array}{rcl}
\logit(\pi_{z;i}) & = & \bsx_{z;i}' \bgamma_{z},
\end{array}
\end{equation}
where $\bsx_{z;i} = (x_{z;i1}, \ldots, x_{z;id_z})$ represents the $i$-th row of the design matrix $\bX_z$ for the logistic regression model and 
$\pi_{z;i} = \pr(Z_i = 1)$ is the probability that $i$-th profile belongs to the variance heterogeneous group.
For the outlier indicators, let $\bX_{w;i} = [x_{w;ijk}]_{jk}$ ($j=1, \ldots, n_i$, $k =1, \ldots, d_w$) be a $n_i\times d_w$ design matrix corresponding to individual $i$ with $x_{w;ijk}$ representing the $k$-th covariate at time $t_{ij}$. 
Note that, we may include time and other relevant time-varying covariates in the design matrix. 
Let $\pr(W_{ij} = 1) = \pi_{w;ij}$ be modeled as
\begin{equation} \label{eq:logistic_w}
\mathrm{logit}(\pi_{w;ij}) = \bsx_{w;ij}'\boldsymbol{\gamma}_w,
\end{equation}
where $\bsx_{w;ij}$ is the $j$-th row of $\bX_{w;i}$ and $\boldsymbol{\gamma}_w$ is the vector of regression coefficients.


\subsection{Prior distribution choices and model identifiability} \label{sec:model_priors}

We now discuss the choice of priors for the model parameters and the scaling parameters $\eta_u, \eta_w,\eta_z$ that determine the relative differences between the homogeneous and heterogeneous groups.  
It is important to realize that our model functionally consists of a collection of mixture model components, and absent informative priors or parameter restrictions, there are many potential configurations of the parameters that can equally fit a particular dataset due to usual label switching considerations \citep{stephensDealingLabelSwitching2000, jasraMarkovChainMonte2005}.  
However, as discussed, we are interpreting the groups defined by the 
active indicators ($U_i=1$, $Z_i=1$, $W_{ij}=1$) to correspond to heterogeneous structures, and thus care must be taken in choosing the prior structures to ensure that the resulting estimates respect this consideration.

First, we consider the parameters $\eta_u$, $\eta_w$ and $\eta_z$.
These variance scaling factors specify how extreme the heterogeneous groups are relative to the corresponding homogeneous groups.
Importantly, each of these scale parameters have similar interpretations, representing the multiplicative increase in the scale of the random effects, the residual (for a single observation), and the residual standard deviation (for all residuals of the individual), respectively.  
The choice of these $\eta$s define what constitutes the heterogeneous clusters, and their magnitude must be made based on user belief of how extreme observations can be while still being understood as consistent with the homogeneous group.  In most cases, a choice of 3, 4, or 5 will represent a reasonable choice such that samples that fall far in the tail of the  dominant homogeneous cluster distribution will be assigned to the heterogeneous cluster.
Our choices for $\eta$s are consistent with those available in the literature of scale contamination models \citep{gleasonUnderstandingElongationScale1993, mccullochFlaggingUnusualClusters2024}.
Given that the hyperparameters operate on the same scale, we generally pick a single choice and set $\eta_u=\eta_w=\eta_z$, although one could make separate choices for each of the components.

While the fixed values of the $\eta$ scaling parameters control the magnitude of the difference between the homogeneous and heterogeneous groups, it is possible for the heterogeneous group to end up with the bulk of observations leading to unreasonable interpretation.  
Consequently, we must also utilize somewhat informative priors on the logistic regression models to ensure that the heterogeneous components do not dominate the mixture.  
To that end, we recall standard results about the asymptotic normality for the estimate of the log-odds.  
For a sample of size $n$ and success rate $p$, an asymptotic distribution for $\logit(\hat{p})$ is N$\left(\logit(p), \frac{1}{n p(1-p)} \right)$, which we use to motivate the choice of a moderately informative prior for the logistic regression intercepts.  
When the covariates within the design matrices for the logistic regression indicators are centered, the intercept $\gamma_{*0}$ ($*=u,w,z$) represents the predicted rate of heterogeneity for a subject/observation at the mean value of all predictors.  
Consequently, we specify the prior as 
\begin{equation} \label{eq:prior_prop}
\gamma_{*0} \sim \text{N} \left(\logit(p_{*}), \frac{1}{n_* p_{*}(1-p_{*})} \right),
\end{equation}
using subjective choices of $p_*$ and $n_*$.  
Generally, we believe that is reasonable to a priori assume that 5\% of the sample is mean heterogeneous, 3\% of the total number of observations are outliers, and 5\% of the sample may be variance heterogeneous; that is, $p_{u} = p_z = 0.05$ and $p_w = 0.03$.  
We use the prior sample size $n_*$ to control how informative this prior will be, and we recommend using a prior with weight equivalent to 20\% of the number of indicators informed by the regression model.  
That is, we use $n_u=n_z=0.2n$ and $n_w=0.2(\sum_{i=1}^n n_i)$.
For the rest of the logistic regression parameters, we choose $\gamma_{u(-0)} \sim \MVN(\bzero, c_u^2 \bI)$, $\gamma_{w(-0)} \sim \MVN(\bzero, c_w^2 \bI)$ and $\gamma_{z(-0)} \sim \MVN(\bzero, c_z^2 \bI)$. 
We choose a relatively tight value such as $c_u = c_w = c_z = 0.1$ to achieve some regularization in the logistic coefficients.

The prior for $\sigma_1^2$, residual variance for the variance heterogeneous group, is chosen such that it is centered around at a value meaningfully larger than $\sigma_0^2$ and is given by
\begin{equation} \label{eq:resid_var_prior2}
\sigma_1^2 | \sigma_{0}^2, \alpha^2 \sim \textrm{IG}\left( \frac{n_\sigma}{\alpha^2}, \left(\frac{n_\sigma}{\alpha^2} - 1 \right) \eta_z^2 \sigma_0^2 \right),
\end{equation}
Note that the shape and rate parameters of the inverse-gamma distribution are chosen such that $\text{E}(\sigma_1^2|\sigma_0^2, \alpha^2) = \eta_z^2 \sigma_0^2$ and $\Var(\sigma_1^2|\sigma_0^2, \alpha^2) = \eta_z^4 \sigma_0^2/(n_\sigma / \alpha^2 -2)$.
Given a set $\{\sigma^2_i\}_{i:Z_i=1}$ and $\sigma^2_0$, the conditional sampling distribution of $\sigma^2_1$ is inverse Gamma with shape parameter 
$\frac{1}{\alpha^2} \left( \sum_{i} Z_i + n_\sigma \right)$
and the scale parameter $\left(\frac{n_\sigma}{\alpha^2}-1 \right) \eta_z^2 \sigma_0^2 + \frac{1}{\alpha^2} \sum_{i:Z_i=1} \sigma_i^2$.  
It is clear from the form of the shape parameter of this sampling distribution that the hyperparameter $n_\sigma$ plays the role of a prior sample size. 
We generally choose $n_\sigma = 0.01n$.
With the previous assumption of 5\% variance heterogeneity, a prior sample size equal to 1\% of the total sample size corresponds to a weight of 20\% of the expected number of variance heterogeneous samples,  consistent with the moderately informative priors we have used for the logistic regression parameters.
Observe from the prior of $\sigma_1^2$ in (\ref{eq:resid_var_prior2}), that we constrain  $\alpha^2 < n_\sigma$ so that scale parameter of this prior is positive.
To that end, we choose the prior for $\alpha^2$ as $\text{Exp}(1)$ truncated at $n_\sigma$.


While the priors corresponding to the heterogeneous model components and their corresponding regression models require moderately informative choices to stabilize inference, the priors corresponding to the fixed effects and other components of the homogeneous model are chosen to be highly disperse and non-informative.
We assume the global variance $\sigma_0^2 \sim \text{IG}(h_1,h_2)$, where $h_1$ and $h_2$ are chosen suitably to obtain a relatively non-informative prior for the overall variance parameter. Generally, we take $h_1=0.1$ and $h_2=0.1$. 
We assume the autoregressive correlation coefficient $\rho \sim \text{Unif}(0,1)$.
The prior of $\bbeta$ is chosen as $\mathrm{MVN}_p \left(\boldsymbol{\mu}_{\beta}, c_{\beta}^2\mathbf{I}_p \right)$. 
Here we choose $\bmu_\beta = \bzero$ and to have a diffuse prior we set $c_{\bbeta} = 10$.
We assume $\bLambda$ to have a prior $\IW(\nu, \bI)$, where $\nu=q+1$.

\subsection{Special nested cases of the HOILD model} \label{sec:model_submodels}

It is important to emphasize that while our  modeling approach provides a framework to simultaneously identify three different sources of heterogeneity, it is flexible and robust enough to continue to model data effectively when all these capabilities may not be necessary.
In particular, a number of special cases  may also be of interest
and can easily be formed by removing some of the mixture components of our strategy.
We refer to our general strategy of combining contamination models for our three sources of heterogeneity as our HOILD modeling approach (Heterogeneity and Outlier Identification in Longidituional Data).  To distinguish special cases where we include or exclude certain contimination components, we provide names for particular choices, including naming our full model as \hemhevo\ to stand for \textit{HE}terogeneous \textit{M}ean, \textit{HE}terogeneous \textit{V}ariance, and \textit{O}utliers.

We consider four additional special cases. 
\begin{enumerate}
    \item 
 $\homhov$ is the standard mixed effects model with homogeneous mean, homogeneous variance and no outliers obtained by fixing $U_i=0$, $W_{ij}=0$, and $Z_i=0$ for all $i,j$.
    \item
 $\hemhov$ is the  model with heterogeneous mean, homogeneous variance, and no outliers, obtained when $U_i$ is random while $W_{ij}$s and $Z_i$s are fixed at zero for all $i,j$. 
    \item 
 $\homhovo$ includes homogeneous mean and variance with outliers, obtained when $U_i = 0$, $V_i = 0$ for all $i$, and $W_{ij}$s are random.
    \item 
 $\hemhovo$ incorporates heterogeneous mean, homogeneous variance and outliers, obtained with random $U_i$ and $W_{ij}$, while $Z_i = 0$ for all $i,j$.
 \item Again, $\hemhevo$ is our full model, where $U_i$, $Z_i$ and $W_{ij}$ are each random Bernoullis.
\end{enumerate}
Many of these special cases of our model are closely related to approaches that have been considered in both frequentist and Bayesian robust-statistics literature.
For instance, \cite{richardsonBoundedInfluenceEstimation1997} presented a frequentist bounded-influence approach whose subject-level weighting structurally resembles the mean heterogeneity contamination structure of our model $\hemhov$.
The independent contamination model of \cite{alqallafPropagationOutliersMultivariate2009} allows for outlier observations, similar to model $\hemhovo$.
\cite{pinheiroEfficientAlgorithmsRobust2001} enables modeling the subject-level mean and variance heterogeneity as in model $\hemhevo$, however, does not allow outlier observations.
The Bayesian contamination frameworks of \cite{bergerRobustBayesEmpirical1986, morenoBayesianRobustnessHierarchical1993}, while not developed for this purpose, could in principle be adapted to model mean heterogeneity as in $\hemhov$.
Heterogeneity in residual variance can also be incorporated into a contamination framework.
However, no existing model simultaneously accommodates longitudinal data with heterogeneous mean, residual variance, and accounts for outlier observations within a single coherent framework -- the gap that our proposed model addresses.

\section{Estimation and Inference} \label{sec:inf}


\subsection{MCMC sampling} \label{sec:inf_mcmc_outline}
We now discuss the MCMC sampling algorithm that is used to perform posterior inference. 
The full details of the MCMC Algorithm is provided in Appendix \ref{appn-sec:comp_algo},
although we comment on a few details here.
We have employed a partially-collapsed Gibbs sampler \citep{vandykPartiallyCollapsedGibbs2008} to sequentially update all model parameters. 
Since many  parameters depend on individual-specific $\sigma_i^2$, we have marginalized those out from the majority of the conditional distributions (sampling steps for $\bbeta$, $\bsb_i$, and $\bLambda$ remain conditional on $\sigma_i^2$).  
This  introduces a blocking structure to improve the mixing of the corresponding Markov chains. 
The logistic regression  coefficients $\bgamma_u$, $\bgamma_w$, $\bgamma_z$ are updated based on a  data augmentation  Gibbs sampling approach proposed by \cite{Polson2013}. 
As suggested by the authors, a set of latent variables  following  P\'{o}lya-Gamma distributions are introduced and the corresponding binomial likelihoods are expressed as a mixture of normal distributions with respect to those latent variables. 

We briefly comment on the sampling of the indicator variables. 
Each subject has an $(n_i+2)$-dimensional vector $(U_i, \bsW_i, Z_i)$  indicating the presence/absence of all heterogeneity components.  Note that sampling this full vector
jointly in every iteration would involve computing $2^{n_i+2}$ probabilities and would be computationally prohibitive.
Conversely, sampling components one-at-a-time conditionally on all others will have poor mixing due to the correlations between the indicators.
To that end, we sample the triple $(U_i, W_{ij}, Z_i)$ conditionally on $\bsW_{i(-j)} = (W_{ij'})_{j' \neq j}$ for each $j=1,\ldots,n_i$.
While updating $U_i$ and $Z_i$ for every update of $W_{ij}$ within the iteration  may initially seem to be computationally inefficient, but it has potential benefits.
Observe that $Z_i$ and $W_{ij}$ are negatively correlated in the sense that a profile with a very small number of extreme observations could be considered as variance homogeneous with outliers or as a variance heterogeneous profile with no outliers and large $\sigma_i^2$.
Furthermore, $Z_i$ is negatively associated with $U_i$. 
A variance hetergeneous profile with $Z_i=1$ will potentially have large $\sigma_i^2$ acounting for bigger fluctuations in the profile, which will in turn discourage the choice of $U_i=1$; in other words, the mean will not be large enough relative to $\sigma_i^2$ to belong to the heterogeneous group.
Therefore, updating $U_i$ and $Z_i$ every time a $W_{ij}$ is updated 
 better facilitates mixing over the joint support of the indicators than updating a single indicator conditional on the others.


\subsection{Fixed effects estimation and heterogeneity identification}
\label{sec:inf_beta_indicators}

The MCMC algorithm is run for a large number of iterations, and the first few iterations are removed until convergence is reached. The remaining samples are thinned and used for posterior inference.
MCMC convergence is verified through Geweke diagnostic tests and we make sure that effective sample sizes are reasonably large (generally, at least 100) for each of the key parameters $\bbeta$, $\bgamma_u$, $\bgamma_w$, $\bgamma_z$, $\hat{u}$, $\hat{w}$, $\hat{z}$, and $\sigma_0^2$.
We determine the fixed effects based on the posterior samples of $\bbeta$, denoted as $\bbeta^{(g)}$, $g = 1, \ldots, G$, where $\bbeta^{(g)} = (\beta_1^{(g)}, \ldots, \beta_p^{(g)})'$. 
Let $\beta_k^{(l)}$ and $\beta_k^{(u)}$ indicate the lower and upper bound of the 95\% highest density interval (HDI) for $\beta_k$. 
We claim that $k$-th predictor is not significant when $\beta_k^{(l)} < 0 < \beta_k^{(u)}$. 
The factors associated with mean heterogeneity, outlier identification and variance heterogeneity will be estimated similarly based on their respective posterior samples denoted by $\bgamma_u^{(g)}$, $\bgamma_w^{(g)}$ and $\bgamma_z^{(g)}$ respectively, while their respective 95\% HDI will be denoted by $(\gamma_u^{(l)}, \gamma_u^{(u)})$, $(\gamma_w^{(l)}, \gamma_w^{(u)})$, and $(\gamma_z^{(l)}, \gamma_z^{(u)})$.

We also aim to identify the profiles which are mean and/or variance heterogeneous and flag the outlier measurements within profiles.
To that end, the indicator variables $U_i$, $Z_i$ and $W_{ij}$ for every individual $i$ and observation $j$ are estimated such that the resulting model has best predictive performance.
\cite{barbieriOptimalPredictiveModel2004} suggest that a median probability model will often be the optimal predictive model.
In our context, this translates to estimating an indicator variable to be 1, when its posterior probability is more than 0.5.



Let $u_i^{(g)}$, $z_i^{(g)}$ and $w_{ij}^{(g)}$, $g=1,\ldots, G$ denote the posterior samples of $U_i$, $Z_i$ and $W_{ij}$ respectively.
The posterior probability that individual $i$ is classified to the mean heterogeneous group is given by $\pr(U_i=1 | \bsy_i) = \Exp(U_i | \bsy_i)$. 
We can estimate this quantity based on the posterior samples as $\widetilde{u}_i = \frac{1}{G} \sum_{g=1}^G u_{i}^{(g)}$. 
The estimated mean heterogeneity identifier for $i$-th individual, denoted by $\widehat{u}_i$ is obtained as $\widehat{u}_i = \ind(\widetilde{u}_i > 0.5)$, where $\ind(\cdot)$ denotes the indicator function.
The overall subject-level heterogeneity in the cohort can then be estimated by $\widehat{u} = \frac{1}{n} \sum_{i=1}^n \widehat{u}_i$.
Similarly we obtain $\widetilde{z}_i$, $\widehat{z}_i$ capturing the subject-level variance heterogeneity for $i$-th individual, and $\widehat{z}$ representing the overall variance heterogeneity in the data.
Identification of the observation-level outliers are captured by equivalent quantities $\widetilde{w}_{ij}$ and $\widehat{w}_{ij}$ and the overall proportion of observation-level outliers is obtained as $\widehat{w} = \frac{1}{\sum_{i=1}^n n_i} \sum_{i=1}^{n_i} \widehat{w}_{ij}$.


\subsection{Selection of model and tuning hyperparameters} \label{sec:inf_model_eta_selection}

Our proposed framework is highly flexible and allows heterogeneity in three different model components.  As noted in Section \ref{sec:model_submodels}, there are a number of meaningful special cases of our model that can be considered by fixing various sets of the heterogeneity-defining indicators at the value of zero.  It may be of interest to fit the full proposed model alongside some of these special cases and perform model selection to determine what choice best fits the data and what types of outliers/heterogeneity are present.

As discussed in the literature \citep{
gelmanUnderstandingPredictiveInformation2014}, 
the Watanabe-Akaike information criterion \citep[WAIC;][]{watanabeAsymptoticEquivalenceBayes2010} 
is one of the most common and best performing metrics for model selection.
The WAIC for a given model is typically computed as
\begin{align} \label{eq:waic}
\WAIC &= -2\left(\sum_{i=1}^n \log \widehat{E}_{\btheta|\bsy} \, p(\bsy_i|\btheta) - p_{\WAIC} \right), \\
p_\WAIC &= 2\sum_{i=1}^n \left(\log \widehat{E}_{\btheta|\bsy} \, p(\bsy_i|\btheta) - \widehat{E}_{\btheta|\bsy} \, \log(p(\bsy_i|\btheta)) \right), \nonumber
\end{align} 
where $p_{\WAIC}$ is the penalty term to adjust for overfitting. 
Note that for any estimand $f(\btheta)$, $E_{\btheta|\bsy}f(\btheta)$ represents the posterior expectation of $f(\btheta)$ over $\btheta$ and is estimated using the posterior samples of $\btheta$ as $\widehat{E}_{\btheta|\bsy}f(\btheta) = \frac{1}{G} \sum_{g=1}^G f(\btheta^{(g)})$, where $\btheta^{(g)}$ is the $g$-th sample drawn from the posterior distribution of $\btheta$. 


In our context,  $\btheta$ indicates the parameter vector $(\bbeta, \bgamma_u, \bgamma_w, \bgamma_z, \sigma_0^2, \sigma_1^2, \alpha, \rho, \bLambda)$, 
and $\btheta$ together with the indicators $u_i$, $\bsw_i$ and $z_i$ represent the full set of parameters.
We first discuss the way of deriving $p(\bsy_i|\btheta)$ by marginalizing $p(\bsy_i|\btheta, u_i, z_i, \bsw_i)$ over all possible values of $u_i$, $z_i$ and $\bsw_i$. 
We obtain $p(\boldsymbol{y}_i | \boldsymbol{\theta}, \boldsymbol{w}_i)$ by marginalizing over $u_i$ and $z_i$ 
(but conditioning on $\bw_i$) as 
\[
p(\bsy_i|\btheta, \bsw_i) = \sum\limits_{u_i=0}^1 \sum\limits_{z_i=0}^1 \pi_{u;i}^{u_i}(1-\pi_{u;i})^{(1-u_i)} \pi_{z;i}^{z_i} (1-\pi_{z;i})^{(1-z_i)} f_{z_i}(\bsy_i | \btheta, \bsw_i),
\]
where $f_{0}(\cdot)$ and $f_{1}(\cdot)$ are described in (\ref{eq:f0_f1_mixture}). 
We let $\mathcal{S}_i$ be the set consisting of all $2^{n_i}$ different possibilities of $\boldsymbol{w}_i$. 
Then $\hat{p}(\boldsymbol{y}_i | \boldsymbol{\theta})$ can be obtained by marginalizing $\boldsymbol{w}_i$ over $\mathcal{S}_i$. 
Since $\mathcal{S}_i$ has large number of elements when $n_i$ is large, it might be more efficient to exclude unlikely choices of $\bsw_i$ from $\mathcal{S}_i$.
Note that $\boldsymbol{w}_i$, being the indicator of outliers for individual $i$, is expected to contain only a small number of ones, 
as an individual subject will not have more than a few measurements classified as outliers;
otherwise it would be a variance heterogeneous patient with an inflated $\sigma^2_i$.
We obtain $\hat{p}(\boldsymbol{y}_i | \boldsymbol{\theta})$ as follows,
\begin{equation}\label{eq:lhd_marginal}
\hat{p}(\boldsymbol{y}_i | \boldsymbol{\theta}) \approx \hat{p}(\boldsymbol{y}_i | \boldsymbol{\theta}, \bsw_i \in \mathcal{T}_i) = \frac{ \sum\limits_{\boldsymbol{w}_i \in \mathcal{T}_i} \left\{\prod\limits_{j=1}^{n_i} \pi_{w;ij}^{w_{ij}} (1-\pi_{w;ij})^{1 - w_{ij}} \right\} p(\boldsymbol{y}_i | \boldsymbol{\theta}, \boldsymbol{w}_i) }{\sum\limits_{\boldsymbol{w}_i \in \mathcal{T}_i} \left\{ \prod\limits_{j=1}^{n_i} \pi_{w;ij}^{w_{ij}} (1-\pi_{w;ij})^{1 - w_{ij}} \right\} },
\end{equation}
where $\mathcal{T}_i$ is an a priori chosen subset of the full set of configurations $\mathcal{S}_i$. 
We consider it to be very unlikely for a profile to have more than two outliers, and therefore restrict our set $\mathcal{T}_i$ to only consist of vectors with up to two ones and zeros for the rest.


We use $\hat{p}(\boldsymbol{y}_i | \boldsymbol{\theta})$ calculated using equation (\ref{eq:lhd_marginal}) as the likelihood for obtaining WAIC in our models.
The best model is selected to be the one with smallest WAIC score.
Since the likelihood used to compute $\WAIC$ are marginalized over the indicators, WAIC measured for the full model as well for all of the special-case nested models will be comparable.
We can also obtain WAIC for our model using the data likelihood conditional on the individual-specific indicators $U_i$, $\bsW_i$ and $Z_i$, as  given in equation (\ref{eq:f0_f1_mixture}).
That is, the indicator terms are basically treated as parameters within the $\btheta$ vector which are averaged in the posterior expectation approximation.
For convenience, we denote the WAIC scores based on the marginalized likelihood (\ref{eq:lhd_marginal}) as M-WAIC, and the statistic based on the conditional likelihood (\ref{eq:f0_f1_mixture}) is denoted by C-WAIC.
It is worth observing that, while the conditional likelihood computation for an individual using (\ref{eq:f0_f1_mixture}) has a complexity $O(1)$, marginal likelihood in (\ref{eq:lhd_marginal}) involves averaging the likelihoods $p(\bsy_i|\btheta, u_i, \bsw_i, z_i)$ over $u_i, z_i \in \{0,1\}$ and $\bsw_i \in \mathcal{T}_i$ resulting in a computation complexity of $O(n_i^2)$.
Hence, C-WAIC can be computed much faster than M-WAIC.



We also consider the choice of the hyperparameters $\eta_u$, $\eta_w$, and $\eta_z$. As discussed in Section \ref{sec:model_priors}, we treat these as fixed tuning parameters and investigate the performance of the model under various choices either to select the ``best'' value of $\eta$ or to assess the sensitivity to the choice of $\eta$.
We primarily recommend choosing $\eta$ based on what level of difference represents a meaningful discrepancy from the population average in the context of the specific application, with a choice of $\eta=3$ representing a reasonable default value. 
However, one can also implement WAIC model selection with a small collection of candidate $\eta$s (possibly with different values for $\eta_u$, $\eta_w$, $\eta_z$) to help inform the appropriate values of the tuning parameters.

R code to implement the MCMC sampling scheme and the computation of the model selection statistics can be found on github at 
\url{https://github.com/anishspace/HOILD/}.


\section{Simulation Studies}\label{sec:simulation}

We use a range of simulation experiments to assess the proposed class of models in terms of their ability to identify the source of heterogeneity, and we also demonstrate their estimation performance for the predictor effects.

\subsection{Data generation strategy}
\label{sec:sim_data_generation}

To construct our simulation study, we use the following data generation strategy.
Each dataset consists of 500 individuals. 
We generate four predictor variables: $X_1$ is a binary variable generated from Bernoulli(0.5), $X_2$ is a continuous variable drawn from a standard normal distribution, and $X_3$ and $X_4$ represent linear ($t$) and quadratic ($t^2$) time effects, respectively.
All four covariates are centered and scaled to have a unit standard deviation.
The outcome trajectory for every individual consists of 10 measurements recorded at $t = 1,\ldots, 10$.
We include all four covariates as fixed effects in the mean model.
In addition to these fixed effects, we consider subject-specific intercept, linear, and quadratic time terms in the random effects design.

We further assume that the mean and variance heterogeneity indicators, $U$ and $Z$, are associated with the baseline values of $X_1$ and $X_2$ and that the outlier indicator is associated with $X_3$ as well $X_1$ and $X_2$, as it could be the case that the profiles have more outliers as time progresses. 
Therefore, the $i$-th row of the design matrices $\bX_u$ and $\bX_z$, denoted as $\bsx_{u;i}$ and $\bsx_{z;i}$, are both given by $\left(1, x_{1;i1}, x_{2;i1}\right)$, consisting of the baseline line values of $X_1$ and $X_2$ for $i$-th individual.
The $j$-th row of the $i$-th design matrix $\bX_{w;i}$, represented as $\bsx_{w;ij}$, is $\left(1, x_{1;ij}, x_{2;ij}, x_{3;ij}\right)$.
We set the true values of $\bgamma_u$ and $\bgamma_z$ such that approximately 5\% of the profiles show mean and variance heterogeneity. 
Additionally, $\bgamma_w$ is chosen so that only 3\% of the total number of observations are outliers, while we make sure that there are at most two outliers per profile.
Outcome values are generated using the true values of all regression coefficients presented in Table \ref{tab:true_regression_params}.

\begin{table}[!tb]
\footnotesize
\centering
\begin{tabular}{cccccc}
\toprule
 Parameters        & Intercept & $X_1$ & $X_2$  & $X_3=t$ & $X_4=t^2$ \\
         \midrule
$\bbeta$     & 5.00      & 2.00 & -1.00 & 0.80   & 0.04                      \\
$\bgamma_u$ & -2.50     & 0.25 & 0.10  & -      & -                         \\
$\bgamma_w$ & -3.50     & 0.50 & -0.20 & 0.10   & -                         \\
$\bgamma_z$ & -3.00     & 0.50 & 0.25  & -      & -                        \\
\bottomrule
\end{tabular}
\caption{\label{tab:true_regression_params}True values of the regression parameters.}
\end{table}

We generate the data using the heterogeneity scaling factors of
$\eta_u = \eta_w = \eta_z = 3$. 
The true values for $\sigma_0^2$ and $\sigma_1^2$ are chosen to be 0.4 and $\eta_z^2 \sigma_0^2 = 3.6$.
For every $i$, we generate the indicator variables $U_i$, $W_{ij}$, and $Z_i$ from $\text{Bern}(\pi_{u;i})$, $\text{Bern}(\pi_{w;ij})$, and $\text{Bern}(\pi_{z;i})$, respectively, where $\pi_{u;i}$, $\pi_{z;i}$ and $\pi_{w;ij}$ are calculated using equations (\ref{eq:logistic_u}--\ref{eq:logistic_w}).
Individual-level residual variances are generated from the distribution 
in (\ref{eq:resid_var_prior}),
where $\alpha^2 = 0.001$.
The random effects $\bsb_i$ and the auto-regressive error $\beps_i$ are generated from the model in (\ref{eq:model_y_eps},\ref{eq:model_mean}).
We choose $\bLambda = ((4.00, 0.05, -0.10), (0.05, 2.25, -0.05), (-0.10, -0.05, 0.64))$ and $\rho = 0.2$.
While generating the data, it is important to ensure that $\bsb_i$ and $\beps_i$ are 
assigned to extreme values when the indicators $U_i$, $\bsW_i$ and $Z_i$ are active.

To ensure that, we use rejection sampling to generate $\bsb_i$ and $\beps_i$ until the following two conditions are satisfied --- 
(a) the set of $\epsilon_{ij}$ corresponding to $W_{ij}=1$ (i.e., outliers) have significantly larger magnitude compared to the non-outliers, that is,
$|\epsilon_{ij}/\sigma_0| > \Phi^{-1}(0.995)$ holds for all $j$ where $W_{ij} = 1$;
(b)
if $U_i=1$ and/or $Z_i=1$, then the likelihood of $\bsY_i = \bX_i \bbeta + \bZ_i\bsb_i + \beps_i$ 
is significantly larger under the true values of the indicator variables $U_i$ and $Z_i$ compared to that under 
subject-level homogeneity
($U_i = Z_i = 0$).
To ensure (b) is satisfied we check that when $u_i+z_i>0$, the following condition is satisfied,
\[
\log p(\bsY_i | U_i=u_i, \bsW_i, Z_i=z_i) - \log p(\bsY_i | U_i = 0, \bsW_i, Z_i = 0) > 2.
\]
If the generated $(\bsb_i,\beps_i)$ fail either (a) or (b), a new set are sampled until that satisfy the conditions for $(U_i,\bsW_i,Z_i)$.


To evaluate the performance in settings where the data include and exclude different types of heterogeneity, our simulation experiments generate data from four different combinations of heterogeneity sources:
(a) $\homhov$: HOmogeneous Mean HOmogeneous Variance without outliers, that is, for each $i$, $U_i=Z_i=0$ and for each $(i,j)$, $W_{ij} = 0$; 
(b) $\hemhov$: HEterogeneous Mean HOmogeneous Variance without outliers, that is, for each $i$, $U_i\in\{0,1\}$, $Z_i = 0$ and for each $(i,j)$, $W_{ij} = 0$; 
(c) $\homhovo$: Homogeneous Mean Homogeneous Variance with Outliers, that is, for each $i$, $U_i=Z_i = 0$ and for each $(i,j)$, $W_{ij} = \{0,1\}$;
(d) $\hemhevo$: HEterogeneous Mean HEterogeneous Variance with Outliers, that is, for each $i$, $U_i \in \{0,1\}$, $Z_i \in \{0,1\}$ and for each $(i,j)$, $W_{ij} \in \{0,1\}$.
We generate 100 datasets for each data setting assuming $\eta_u = \eta_w = \eta_z = \eta = 3$.
We fit all five HOILD models ($\homhov$, $\hemhov$, $\homhovo$, $\hemhovo$, $\hemhevo$) using the true value of $\eta$.
To assess the sensitivity of our choice of $\eta$ to fixed effects estimation and heterogeneity identification, we also fit the models using $\eta$ from $\{2,4,5\}$.
See Appendix \ref{appn-sec:eta} for the $\eta$ sensitivity analysis.

In addition to our proposed modeling framework, we also 
study competitor models based $t$-distribution assumptions that are typically used to model data with extreme observations \citep{welsh13ApproachesRobust1997, pinheiroEfficientAlgorithmsRobust2001}.
We considered three such models by tweaking the distributions for random effects and residuals:
(1) $\tR$: assume $b_i \sim \MVt_5(\bzero, \sigma_0^2 \bLambda)$ and $\beps_i \sim \MVN(\bzero, \sigma_0^2 \Omega(\rho))$ in equation (\ref{eq:model_y_eps}) to generate trajectories with extreme means and normally-distribution residual errors (no outliers), 
(2) $\tE$: assume $b_i \sim \MVN(\bzero, \sigma_0^2 \bLambda)$ and $\beps_i \sim \MVt_5(\bzero, \sigma_0^2 \bOmega)$ that allows generating trajectories with extreme residuals and mean homogeneity, and 
(3) $\tRE$: assume $b_i \sim \MVt_5(\bzero, \sigma_0^2 \bLambda)$ and $\beps_i \sim \MVt_5(\bzero, \sigma_0^2 \bOmega)$ allowing both types of extreme behavior in the data. 
We use these models to fit the data generated under our proposed structure to investigate the benefit of our methods relative to competitors in settings most favorable to our assumptions.  
Additionally we generate data under each of these three choices to assess the performance of our approach in a less favorable scenario (see Appendix \ref{appn-sec:fixed_effects_t}).

We run the MCMC algorithm for every dataset and model pair for $G=6000$ iterations, and after removing first 2000 samples as burn-in and considering every 4th from the rest, we obtain 1000 posterior samples in each case.
In our implementation, the standard mixed effects model $\homhov$ runs in under less than 3 minutes, whereas the most complex model $\hemhevo$ needs around 16 minutes
to generate posterior samples.


\subsection{Model selection}

We first evaluate model selection using datasets simulated from our proposed framework alongside $t$-distribution based models (\tR, \tE, and \tRE). 
For the proposed models, each are fit using $\eta = 3$, which corresponds to the ground truth used during data generation.
For every model-data configuration we evaluate both M-WAIC and C-WAIC, and compute the proportion of times each model is selected under each data setting.
Table \ref{tab:sim_model_selection} shows that both M-WAIC and C-WAIC perform very well as model selection criteria, with M-WAIC only slightly better. 
When data are generated from one of our HOILD models, WAIC selects the correct generating model almost perfectly. The exception is that there is with some tendency to include mean heterogeneity in the setting where data are generated with only outliers ($\homhovo$).

To consider performance when data are generated from a non-HOILD model, we see that when data come from the thick-tailed random effects truth ($\tR$), $\hemhov$ with mean-heterogeneity was selected 100\% of the times by both C-WAIC and M-WAIC. 
This is clearly  the HOILD model that best matches the data generation scheme.
For $\tE$ with thick-tailed residuals, WAIC consistently selects $\hemhevo$. 
Note that $\tE$ assume the full residual vector comes from a $t$-distribution, not the individual components; this is more consistent with our thick-tailed marginal GAL distribution  (\ref{eq:f0_f1_mixture}) corresponding to observations with $Z_i=1$ and variance heterogeneity.  
Hence, $\hemhevo$ is the closest model structure to the truth among our five versions.  
For the same reasons, $\tRE$ with both thick-tailed random effects and residual vectors consistently selects $\hemhevo$ as the preferred model.
These results demonstrate the efficacy of our model selection strategy, while also highlighting that the computationally efficient alternative C-WAIC is almost as effective as M-WAIC.


Alongside the ability of WAIC to select between the competing heterogeneous models, making a ``good'' choice for the tuning parameter $\eta$ is important.
Since $\eta$ determines the extremeness of the heterogeneous group and the observation-level outliers, its selection can be subjective; a clinician might suggest a value tailored to the specific data at hand.
However, a formal model selection strategy can be a useful supplement to guide this process.
To that end, we provide a thorough discussion on model selection for $\eta$ in Appendix \ref{appn-sec:eta_model_selection}.
We observe that both WAIC metrics generally perform well at recovering the true $\eta=3$ relative to $\eta\in\{2,4,5\}$, with M-WAIC demonstrating superior performance to C-WAIC.



\begin{table}[!tb]
    \centering
    \footnotesize
    \begin{tabular}{rccccc}
        \toprule
        & $\homhov$                & $\hemhov$  & $\homhovo$              & $\hemhovo$              & $\hemhevo$ \\ 
        \midrule
        Data Generating Model: & \multicolumn{5}{c}{Proportion C-WAIC Selected Model} \\
        \addlinespace
        \hspace{1em}$\homhov$   & \bf{0.91} &   0.09 & 0.00                  & 0.00                   & 0.00                 \\
        \hspace{1em}$\hemhov$   & 0.00                   & \bf{1.00} & 0.00 & 0.00                   & 0.00  \\
        \hspace{1em}$\homhovo$ & 0.00                   & 0.00     & {\bf 0.75}               & 0.24 & 0.01 \\
        \hspace{1em}$\hemhevo$ & 0.00                   & 0.00     & 0.00               & 0.00  & \bf{1.00} \\ 
        \hspace{1em}$\tR$ & 0.00                   & {\bf 1.00}     & 0.00             & 0.00 & 0.00   \\ 
        \hspace{1em}$\tE$ & 0.00                   & 0.00     & 0.03              & 0.01 & \bf{0.96}  \\ 
        \hspace{1em}$\tRE$ & 0.00                   & 0.00    & 0.00               & 0.02 & {\bf 0.98} \\ 
        
        \midrule
        & \multicolumn{5}{c}{Proportion M-WAIC Selected Model}                \\
        \addlinespace
        \hspace{1em}$\homhov$      & \bf{1.00} & 0.00   &  0.00      & 0.00      & 0.00                   \\
        \hspace{1em}$\hemhov$   & 0.00 & \bf{1.00} & 0.00 & 0.00  & 0.00                 \\
        \hspace{1em}$\homhovo$ & 0.00                   & 0.00     & {\bf 0.64}               & 0.35  & 0.01 \\  
        \hspace{1em}$\hemhevo$ & 0.00                   & 0.00     & 0.00               & 0.00 & \bf{1.00} \\ 
        \hspace{1em}$\tR$ & 0.00                   & {\bf 1.00}     & 0.00             & 0.00 & 0.00 \\ 
        \hspace{1em}$\tE$ & 0.01                   & 0.00     & 0.00               & 0.00 & \bf{0.99} \\ 
        \hspace{1em}$\tRE$ & 0.00                   & 0.00    & 0.00               & 0.00 & {\bf 1.00}  \\ 
        
        \bottomrule
		\end{tabular}
		\caption{WAIC based model selection under three different data settings. Values represent the proportion of generated data sets in which the data analysis model (column) is selected as the best model; values in bold designate the model selected most often for the data generating model (row).}
		\label{tab:sim_model_selection}
\end{table}

\subsection{Fixed effects inference} \label{sec:sim_fixed_eff}

We compare the estimation of fixed effects across different models under four data settings with varying heterogeneity types using three performance metrics.
For a particular pair of data setting and model choice, let $\hat{\beta}_{k,b}$ represent the estimate of $k$-th component of $\bbeta$ obtained by fitting the model to the $b$-th dataset where $b=1,\ldots,B$, and we let the corresponding true value be denoted by $\beta_{k,0}$.
We denote the corresponding 95\% credible interval by $(\beta_{k,b}^{(l)}, \beta_{k,b}^{(u)})$.
To assess the estimation performance, we compute the sum of squared error (SSE) for $\bbeta$ for $b$-th dataset as $SSE_b=\sum_{k=1}^p (\beta_{k,b} - \beta_{k,0})^2$.
The 95\% coverage rate for $\beta_k$ is computed as $\frac{1}{B}\sum_{b=1}^B \ind \left( \beta_{k,0} \in (\beta_{k,b}^{(l)}, \beta_{k,b}^{(u)}) \right)$.
Thus for each model and data settings, we obtain $p$ coverage rates.
We calculate the 95\% binary confidence interval for each of these coverage rates following \cite{agrestiApproximateBetterExact1998}, and compare the intervals against the nominal 95\% target.
To evaluate the efficiency gain we achieve in $\bbeta$ estimation by accounting for heterogeneity, it is useful to compare the 95\% credible interval lengths for different heterogeneous models against that obtained under the usual random effects modeling choice (full homogeneity).
More efficient estimation will lead to narrower credible intervals.
We define the relative interval length (RIL) for the $k$-th component of $\bbeta$ and $b$-th dataset as $\RIL_{k,b} = \left(\beta_{k,b}^{(u)} - \beta_{k,b}^{(l)} \right) / \left( \tilde{\beta}_{k,b}^{(u)} - \tilde{\beta}_{k,b}^{(l)} \right)$ where $(\beta_{k,b}^{(l)}, \beta_{k,b}^{(u)})$ and $(\tilde{\beta}_{k,b}^{(l)}, \tilde{\beta}_{k,b}^{(u)})$ represent the 95\% credible interval under the target model and the standard mixed effects model $\homhov$, respectively.

\begin{figure}[!tb]
\centering
\includegraphics[width=0.9\linewidth]{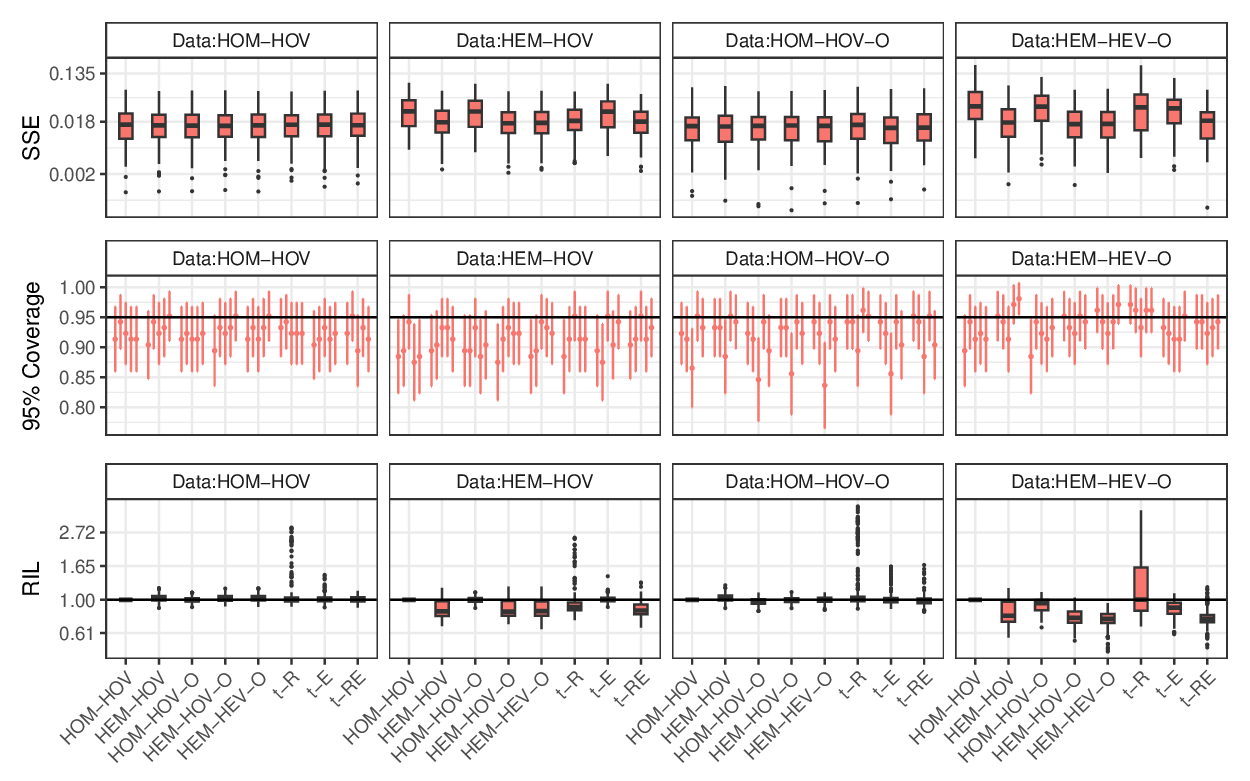}
 \caption{Fixed effects estimation performance across different models in three different data settings.
 Each column in the plot corresponds to model performances under a particular data setting.
 The boxplots in the first (SSE) row display the distribution of SSE evaluated across the 100 simulations for each candidate model fit.
 In the second row, five vertical lines for each model corresponds to the five $\beta$ parameters. 
 Each line represents the 95\% binary confidence interval for a specific $\beta$, estimated from its empirical 95\% coverage rate.
 Third row shows the distribution of the 5x100 RILs in each data and model setting.
 }
\label{fig:sim_fixed_eff_estimation}
\end{figure}

Figure \ref{fig:sim_fixed_eff_estimation} shows the boxplots comparing the estimation performance of all models under different heterogeneity types.
We observe that the SSE for $\bbeta$ are very similar across all models when the data are generated from the standard random effects model.
However, for data generated with mean heterogeneity ($\hemhov$), $\bbeta$ have considerably larger SSE for the fitted models $\homhov$, $\homhovo$, and $\tE$ which assume homogeneous mean, compared to the other models that do not. 
This suggests that ignoring heterogeneity leads to less accurate estimation of $\bbeta$.
Due to a very limited amount of extreme observations in the datasets generated under the outlier-only heterogeneous model $\homhovo$, the SSE for all the models appeared to be very similar.
Like $\hemhov$, the estimated $\bbeta$ for data generating model $\hemhevo$ has largest SSE for model $\homhov$.
While the true model $\hemhevo$ performs best, the SSE for $\hemhov$, $\hemhovo$, and $\tRE$ are very similar.

Under the homogeneous data setting, all models appear to provide reasonable coverage for the $\bbeta$ parameters.
For the datasets generated from $\hemhov$, all models that allow mean heterogeneity provide better coverage of $\bbeta$ compared to those that do not, with $\homhov$ and $\homhovo$ being particularly worse.
When the datasets only have few observation-level outliers ($\homhovo$), however, the coverage for $\bbeta$ turned out to be similar across all models.
In contrast, for the generated data including all sources of heterogeneity, the models $\hemhov$ and $\tR$ appear to slightly overcover $\bbeta$ due to probably having wider credible intervals.
Models $\hemhovo$, $\hemhevo$, and $\tRE$ perform well in this case.

The relative interval lengths of the 95\% credible intervals across all parameters in $\bbeta$ are very similar for every model when data generating model is $\homhov$.
This implies that we do not incur a cost using more complex models than necessary on the homogeneous datasets, in the sense that we still obtain accurate $\bbeta$ estimation with approximately similar credible interval lengths. 
In contrast, the $t$-distribution models occasionally yield substantially less efficient estimation.
When the generated data are heterogeneous, our models allowing heterogeneity provide narrower 95\% credible intervals than $\homhov$,
even as they have slightly higher coverage rates.
This is most prominent when $\hemhevo$ is the data generating model.
In this case, $\hemhevo$ performs best, closely followed by $\tRE$,
while $\hemhov$, $\hemhovo$, and $\tE$ have wider intervals.

We note here that we are investigating the performance of the fixed effects estimation  under the assumption that the true $\eta$ is being used. The sensitivity in coefficient estimation to the choice of $\eta$ is investigated in Appendix \ref{appn-sec:eta_fixed_effects}, where we observe similar SSE and coverage for $\bbeta$ across all $\eta$, with some differences in RIL.
Additionally, we compare $\bbeta$ estimation performance across models under $t$-distributed data generation in Appendix \ref{appn-sec:fixed_effects_t}.



\subsection{Heterogeneity identification}

\begin{figure}[!tb]
\centering
\includegraphics[width=0.7\linewidth]{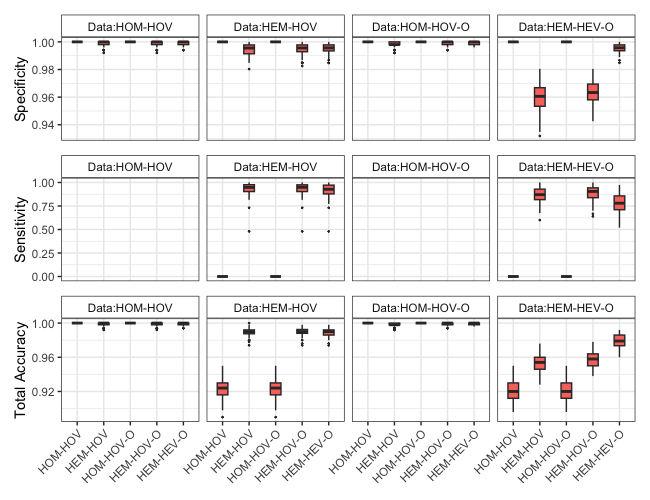}
\caption{The specificity, sensitivity and total accuracy of the subject-level mean heterogeneity indicator $U$ across different models under different data settings.
}
\label{fig:sim_u_estimation}
\end{figure}

We now compare our five proposed models in terms of their ability to identify heterogeneity and outliers under different data scenarios.
As the $t$-distribution models do not make a binary decision regarding whether a data point is consistent with the dominant homogeneous behavior versus heterogeneity, we do not consider their performance in this analysis.
We compare the performance of our models in terms of their specificity, sensitivity and total accuracy to identify each of the three different heterogeneity types.
Specificity of a model to identify a source of heterogeneity is determined by its ability to correctly identify the truly homogeneous profiles, while the sensitivity shows its performance in identifying truly heterogeneous profiles.
Total accuracy of a model provides the overall rate of correct identification.

Figure \ref{fig:sim_u_estimation} compares the performance of all models based on their mean heterogeneity identification by comparing the estimated heterogeneity indicator $\widehat{u}_i$ and the true $U_i$.
Model $\homhov$ and $\homhovo$ do not account for mean heterogeneity, and therefore have $\widehat{u}_i = 0$ for every $i$ resulting in 100\% specificity irrespective of the data generating model.
Since the other models account for heterogeneity and/or outliers, some of the homogeneous profiles will be wrongly classified as heterogeneous resulting in inferior specificity.
When the generative model $\hemhov$ introduces mean heterogeneity in the data, all the models that accommodate mean heterogeneity ($\hemhov$, $\hemhovo$, and $\hemhevo$) perform comparably achieving higher than 99\% specificity.
Among the data generated using the outlier only model $\homhovo$, we observe close to 100\% specificity across all models that allow for mean heterogeneity.
Regarding the generative model $\hemhevo$, the true model provides the best performance among all the models that introduce heterogeneity, while the specificity across all models is  more than 95\% in most datasets.
Overall, all models provide very good performance in terms of identifying homogeneous profiles under every data setting.

The second row in Figure \ref{fig:sim_u_estimation} compares the sensitivity to identify mean heterogeneity between models.
The data generated from $\homhov$ and $\homhovo$ do not have any truly heterogeneous profiles, and therefore the first panel is empty. 
When the data have mean heterogeneity, $\homhov$ and $\homhovo$ being mean homogeneous models do not identify any profile as heterogeneous, corresponding to sensitivity of zero.
All heterogeneous models behave very similarly, with more than 80\% sensitivity for most datasets.
$\hemhovo$ performs best in terms of sensitivity, followed by $\hemhov$ when the true data generative model is $\hemhevo$.
Since $\hemhevo$ accommodates all sources of heterogeneity, it has higher chance of missclassification by identifying a different source of the heterogeniety. 
Note that the variance of the random effects involves $\sigma_i^2$ as well as $(\eta_u^{2})^{u_i}$.
As $U_i$ and $Z_i$ are related, a mean heterogeneous, variance homogeneous profile may be incorrectly classified as a mean homogeneous, variance heterogeneous profile.
In terms of total accuracy for identifying mean heterogeneity, as shown in the third row of Figure \ref{fig:sim_u_estimation}, the true generative models show the best performance in their respective datasets, with the full $\hemhevo$ performing well in all cases even it is more complex than the generating model.

\begin{figure}[!tb]
\centering
\includegraphics[width=0.7\linewidth]{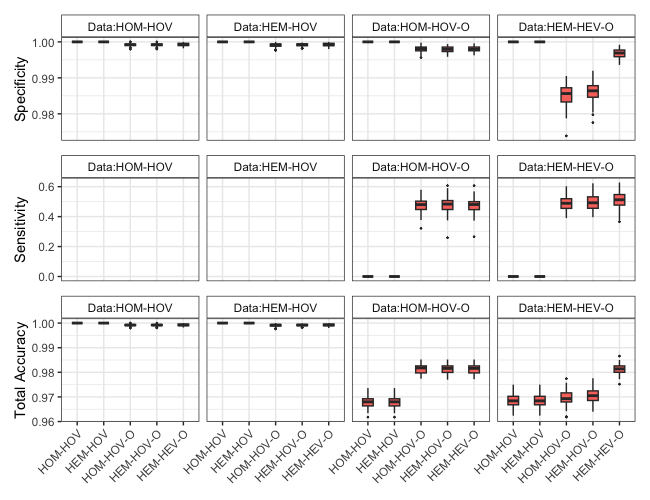}
\caption{The specificity, sensitivity and total accuracy of the observation-level outlier indicator $W$ across different models under different data settings.}
\label{fig:sim_w_estimation}
\end{figure}

Figure \ref{fig:sim_w_estimation} provides comparison of outlier identification performance across all models.
For data model $\homhov$, the no-outlier models $\homhov$ and $\hemhov$
have 100\% specificity, while $\homhovo$, $\hemhovo$, and $\hemhevo$ 
accurately classify almost all non-outlying observations.
Almost identical specificity is observed across all models when the data generative model is $\hemhov$.
For data generated from $\homhovo$, all models accounting for observation-level outliers have higher than 99\% specificity.
In scenarios where all three sources of heterogeneity are present in the generated data ($\hemhevo$), superior performance of $\hemhevo$ is observed relative to $\homhovo$ and $\hemhovo$.
As the latter do not account for variance heterogeneity, it may try to explain higher residual variance in some of the variance heterogeneous, no outlier profiles by considering some of the observations as outliers resulting in lower specificity.
Overall, all models across all data settings provide very high (>98\%) specificity.

The first two panels in the second row do not show any sensitivity results as the corresponding data settings do not incorporate outlier observations in the generated datasets.
Since the models $\homhov$ and $\hemhov$ assume the absence of outliers in the data, they do not provide outlier identification. 
The models $\homhovo$ and $\hemhovo$ show sensitivity comparable to the true model $\hemhevo$. 
The sensitivity to identify observation outliers in all models is close to 50\%.
In terms of total accuracy, the true generative models provide the best performance with at least 98\% total accuracy.

\begin{figure}[!tb]
\centering
\includegraphics[width=0.7\linewidth]{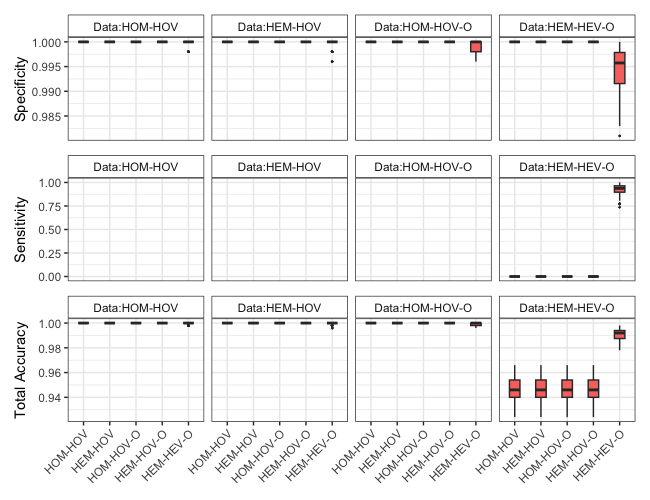}
\caption{The specificity, sensitivity and total accuracy of the subject-level variance heterogeneity indicator $Z$ across different models under different data settings.}
\label{fig:sim_z_estimation}
\end{figure}

The identification performance of different models for variance heterogeneity is accessed under different data settings in Figure \ref{fig:sim_z_estimation}.
$\hemhevo$ is the only model considered that accounts for variance heterogeneity, while the rest of the models identify all the profiles as homogeneous.
Hence, the models $\homhov$, $\hemhov$, $\homhovo$, and $\hemhovo$ have 100\% specificity in all data settings, while $\hemhevo$ also identifies variance homogeneous profiles with very high accuracy of more than 99\%.
The first three panels in the second row of Figure \ref{fig:sim_z_estimation} are empty, as the corresponding data generative models do not allow for variance heterogeneity.
The fourth panel shows that $\hemhevo$ has close to 90\% sensitivity.
All models have very high total accuracy with $\hemhevo$ correctly identifying variance homogeneity/heterogeneity for more than 98\% of the profiles.

All the results discussed so far are based fitting the models using the same value of $\eta$ as was used in the data generation.
We refer the reader to Appendix \ref{appn-sec:eta_heterogeneity} where we show that  identification of heterogeneity is generally robust  to the $\eta$ choice.

We conclude this section by summarizing our findings.
We observe efficiency gains with our proposed model in the estimation of fixed effects, compared to the standard mixed effects model, when the data has different sources of heterogeneity.
Even when the data is homogeneous, the additional complexity of the model does not affect the parameter estimation.
Our proposed model also provides very high specificity and reasonably good sensitivity in identifying the subject-level mean and variance heterogeneity, along with observation-level outliers.
Therefore, using the full model as a default choice can be easily recommended in most scenarios, while the nested models $\hemhov$ and $\hemhovo$ can also be useful if suggested by the WAIC based model selection approach.


\section{Analysis of Heterogeniety and Outliers in Longitudinal Hormome Data}
\label{sec:real_data}

We now demonstrate the utility of HOILD models using data from the well-known Study of Women's Health Across the Nation (SWAN) --- a multi-site prospective study which began in 1994 and continued through 2007.
We refer the reader to \cite{sutton-tyrrellStudyWomensHealth2010, sutton-tyrrellStudyWomensHealth2014}
for additional details on the data.
A central component of the hormonal profile recorded in SWAN is {\it DeHydroEpiAndrosterone Sulfate} (DHEAS). 
It is the primary adrenal steroid biomarker that acts as a unique dual-process indicator tracking both chronological aging and the menopausal transition.
\cite{ghebreAssociationDHEASBone2011} links characteristic fluctuations and increased DHEAS levels during the late menopausal transition with depressive symptoms and anxiety, while \cite{morrisonHigherDHEASDehydroepiandrosterone2011} highlights its role as a key cardiovascular and musculoskeletal marker.
To accurately model the DHEAS trajectories, it is essential to account for presence of individuals with unrepresentative trajectories, extreme values at individual time points, and varying levels of variability to obtain trustworthy inference on population-average effects.

\begin{figure}[!tb]
  \centering
  \includegraphics[width=0.7\textwidth]{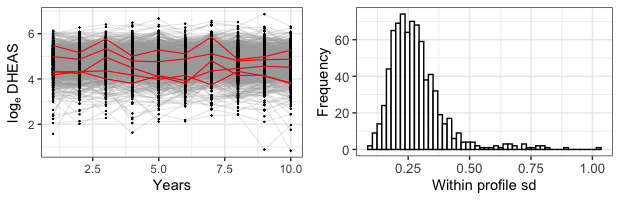}
  \caption{Overview of the log-transformed DHEAS concentration data.
  The trajectories of 5 randomly selected patients are highlighted in red.
  }
  \label{fig:DHEAS}
\end{figure}

We considered yearly DHEAS measurements from 778 study participants as the longitudinal outcome in our analysis. 
Since the measurements were highly skewed, we first log-transformed the concentrations. 
For the mean model, the potential predictors included: (1) a cubic function of the time (in years) denoted as $t$, where $t = 1, \ldots, 10$, that represents the visit number with $t=1$ representing the baseline, along with (2) age at baseline and (3) time-varying BMI.
All continuous covariates were centered and scaled.
Random effects consisted of a subject-specific intercept and a linear slope ($t$).
Furthermore, we considered age and BMI at baseline as relevant predictors influencing subject-level mean and variance heterogeneity. 
To model whether a DHEAS measurement could be considered an outlier, we have chosen $t$, age at baseline and BMI at baseline as the associated factors.
Figure \ref{fig:DHEAS} shows that the distribution of the within-profile standard deviation for all subjects; the long tail in this distribution may indicate the presence of some sources of heterogeneity.


We fit the data using the five version of our HOILD model previously considered.
The same set of covariates are considered across all models for the fixed effects.
The sets of covariates used to model the relevant heterogeneity indicators were also same for all these models. 
We chose $\eta_u = \eta_w = \eta_z \in \{3, 4, 5\}$.

\begin{table}[!h]
\centering
\footnotesize
\begin{tabular}[t]{lccccccc}
\toprule
Model($\eta_u, \eta_w, \eta_z$) & M-WAIC & C-WAIC & $\widehat{u}$ & $\widehat{w}$ & $\widehat{z}$ & $\widehat{\sigma_0^2}$ & 95\% CI \\
\midrule
$\homhov$ & 11914.4 & 11914.4 & 0 & 0 & 0 & 0.084 & (0.081, 0.087)\\
\addlinespace
$\hemhov$ (3,-,-) & 11742.7 & 11551.8 & 0.032 & 0 & 0 & 0.081 & (0.078, 0.084)\\
$\hemhov$ (4,-,-) & 11748.9 & 11555.8 & 0.024 & 0 & 0 & 0.081 & (0.079, 0.084)\\
$\hemhov$ (5,-,-) & 11762.6 & 11571.6 & 0.022 & 0 & 0 & 0.082 & (0.079, 0.085)\\
\addlinespace
$\homhovo$ (-,3,-) & 11085.1 & 9699.2 & 0 & 0.016 & 0 & 0.061 & (0.057, 0.065)\\
$\homhovo$ (-,4,-) & 11628.0 & 9688.6 & 0 & 0.013 & 0 & 0.062 & (0.059, 0.066)\\
$\homhovo$ (-,5,-) & 12181.0 & 9749.9 & 0 & 0.012 & 0 & 0.063 & (0.059, 0.066)\\
\addlinespace
$\hemhovo$ (3,3,-) & 10987.1 & 9567.0 & 0.013 & 0.015 & 0 & 0.060 & (0.056, 0.064)\\
$\hemhovo$ (4,4,-) & 11527.1 & 9534.8 & 0.012 & 0.013 & 0 & 0.061 & (0.057, 0.064)\\
$\hemhovo$ (5,5,-) & 12090.8 & 9598.1 & 0.009 & 0.012 & 0 & 0.062 & (0.059, 0.065)\\
\addlinespace
$\hemhevo$ (3,3,3) & 10881.3 & 9633.9 & 0.004 & 0.012 & 0.022 & 0.058 & (0.054, 0.061)\\
$\hemhevo$ (4,4,4) & 11461.3 & 9672.2 & 0.003 & 0.010 & 0.019 & 0.059 & (0.055, 0.062)\\
$\hemhevo$ (5,5,5) & 12028.4 & 9723.3 & 0.001 & 0.010 & 0.018 & 0.060 & (0.057, 0.064)\\
\bottomrule
\end{tabular}
\caption{Estimation performance of different models for DHEAS data with different $\eta$'s in terms of WAIC, rates of three heterogeneity types, and the homogeneous residual variance $\sigma_0^2$.
}
\label{tab:DHEAS_model_selection}
\end{table}

From Table \ref{tab:DHEAS_model_selection}, we see that the $\homhov$ model that does not account for the heterogeneity and outliers in the data yields the largest/worst WAIC values.  To understand how the inclusion of various combinations of the contamintation components impacts estimation, we include the estimate and CI for the residual variance of each model; for the variance heterogeneous models, we use $\sigma_0^2$.
The $\hemhov$ model that only accommodates mean heterogeneity, but does not account for the outliers and variance heterogeneity,  behaves almost equally worse with slightly decreased estimate for $\sigma_0^2$. This model estimates around 3\% of the 778 individuals to be  the mean heterogeneous group, with the proportion decreasing as $\eta_u$ increases.
The inclusion of outliers either without mean heterogeneity ($\homhovo$) or with mean heterogeneity ($\hemhevo$) yields a sharp drop in $\hat{\sigma}_0^2$ and improved WAIC for $\eta=3,4$.
These models identify around 1\% outliers among all the 7780 measurements.
However, the lowest M-WAIC and the selected model is taken to be $\hemhevo$, which identifies 0.4\% of individuals to be mean heterogeneous and 2.2\% to be variance heterogeneous, with 1.2\% of observations to be outliers.

\begin{table}[!tb]
\footnotesize
\centering
\begin{tabular}{lccccccc}
\toprule
           &  & \multicolumn{2}{c}{$\homhov$}           &  & \multicolumn{3}{c}{$\hemhevo$}                \\
              &  & \multicolumn{2}{c}{(mixed effects model as reference)}           &  & \multicolumn{3}{c}{(full heterogeneous selected model)}                \\
Parameters &  & $\widehat{\bbeta}$ & 95\% CI          &  & $\widehat{\bbeta}$ & 95\% CI          & RIL  \\ \midrule
Intercept & & 4.727 & (4.688, 4.763) & & 4.760 & (4.722, 4.796) & 0.99\\
t & & 0.051 & (-0.041, 0.149) & & 0.049 & (-0.039, 0.132) & 0.90\\
t$^2$ & & 0.020 & (-0.182, 0.228) & & -0.002 & (-0.202, 0.164) & 0.89\\
t$^3$ & & -0.022 & (-0.138, 0.111) & & -0.002 & (-0.114, 0.104) & 0.88\\
Age & & -0.071 & (-0.128, -0.021) & & -0.065 & (-0.119, -0.017) & 0.96\\
BMI & & -0.086 & (-0.113, -0.055) & & -0.081 & (-0.110, -0.055) & 0.95\\
\bottomrule
\end{tabular}
\caption{The table presents the fixed effects of the predictors. Relative interval length (RIL) represents the ratio of the interval length of $\hemhevo$ to $\homhov$.
}
\label{tab:DHEAS_preds_effect}
\end{table}

The estimated coefficients of the fixed effects predictors for the models $\homhov$ and $\hemhevo$ with $\eta_u = \eta_w = \eta_z = 3$ are summarized in Table \ref{tab:DHEAS_preds_effect}. 
The coefficients are approximately the same between the models, with slight variations in the quadratic and cubic time effects.
Under both models, DHEAS measurements decrease significantly with age at baseline and BMI.
The time variable does not appear to have a significant fixed effect, implying that the log-DHEAS concentrations do not have an overall temporal trend. 
It is important to note that by accounting for heterogeneity in the model, the fixed effects for $\hemhevo$ have between 4\% and 12\% narrower credible intervals, compared to the CIs obtained from the model $\homhov$; this  indicates increased precision is achieved under our proposed methodology.

\begin{figure}[t]
    \centering
    \includegraphics[width=0.7\textwidth]{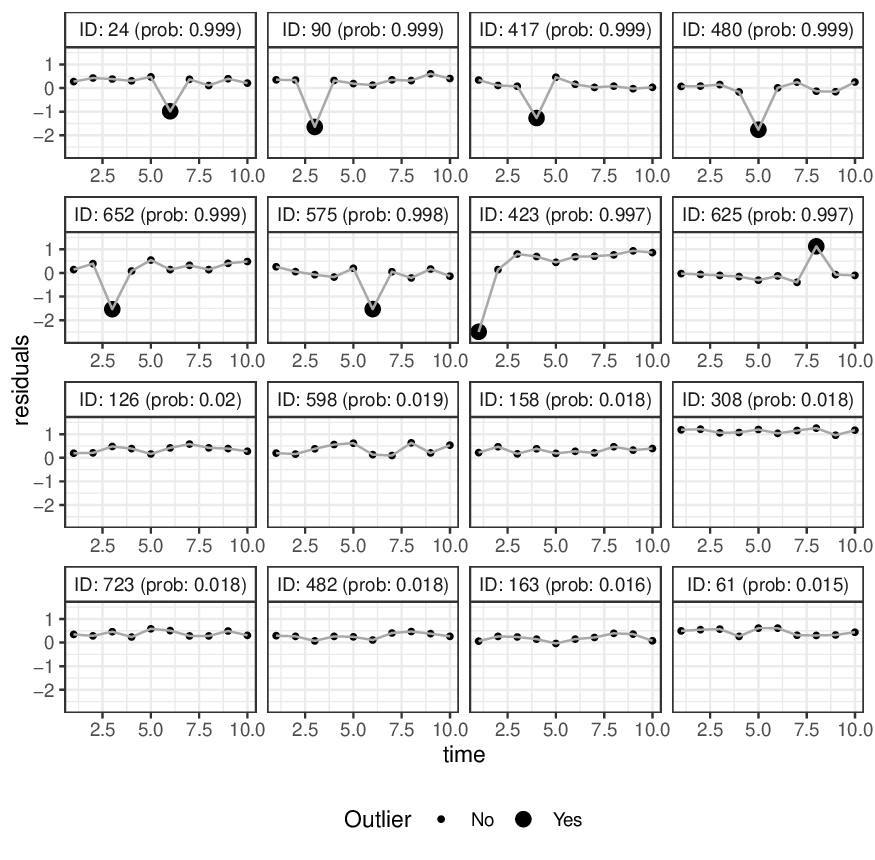}
    \caption{Fixed effects residual profiles ($\hat{\beps}_i = \bsy_i - \bX_i\hat{\bbeta}$) with larger dots representing observations identified as outliers. 
    The headers report the maximum $\widetilde{w}_{ij}$ for the profile. The profiles are in decreasing order of the probabilities, the top-left subfigure displaying the trajectory of subject id 90, whose DHEAS concentration at $t=3$ has the highest probability of being an outlier among all the outcome values across all the subjects.}
    \label{fig:DHEAS_obs_outliers}
\end{figure}
 
We now consider the identification of observation level outliers and the subject level mean and variance heterogeneity in the dataset.
The first two rows in Figure \ref{fig:DHEAS_obs_outliers} show eight subject profiles with the measurements that have the highest posterior probability of being outliers,
along with eight other profiles without any probable outliers.
It is evident that the observations marked as outliers do not fit well with the rest of the profile.
For example, the individual with ID: 417, has its fourth observation marked as an outlier with a probability of approximately 0.999.
No outlier observation is identified within the profiles in the last two rows, since for each of them the largest $\widetilde{w}_{ij}$ is less than or equal to 0.02.

\begin{figure}[!tb]
    \centering
    \includegraphics[width=0.7\textwidth]{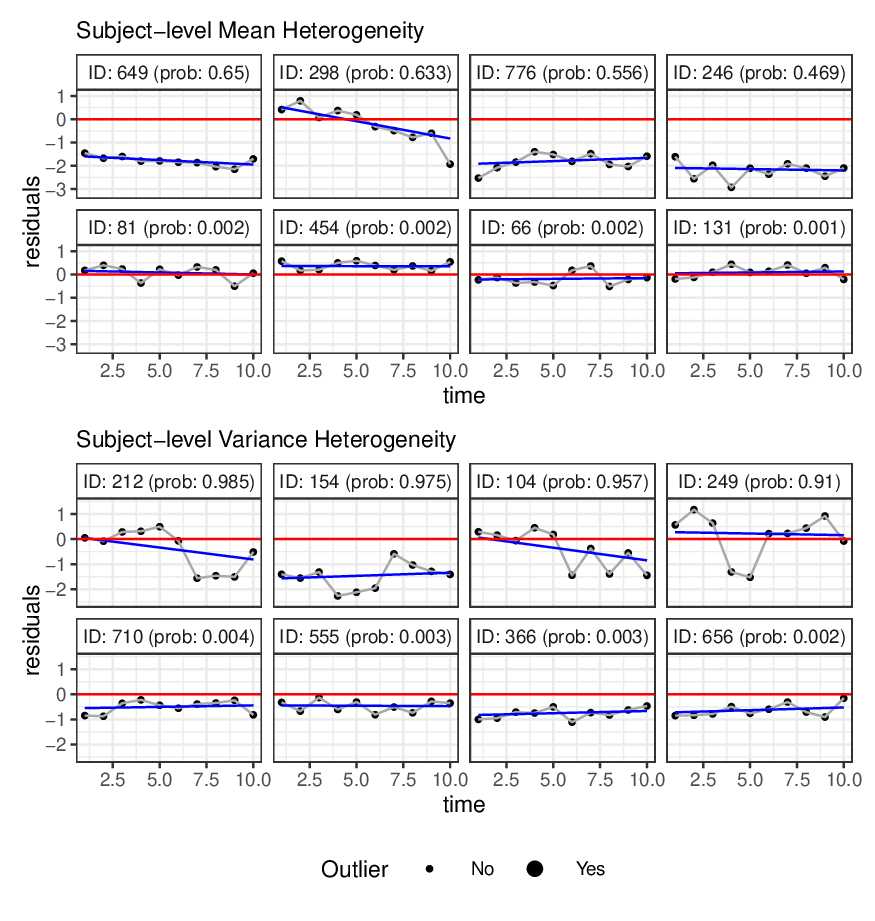}
    \caption{Fixed effects residual profiles ($\hat{\beps}_i = \bsy_i - \bX_i\hat{\bbeta}$) with subject-level mean and variance heterogeneity. 
    First two rows show profiles with most and least probable mean heterogeneous profiles respectively and the bottom two rows show profiles with most and least variance heterogeneity.
    The blue line shows the random effects $\bZ_i\bsb_i$. 
    The probabilities for the relevant heterogeneity are reported in the header of each subfigure.
    }
    \label{fig:DHEAS_mean_var_het}
\end{figure}

The subject-level mean heterogeneities are shown in Figure \ref{fig:DHEAS_mean_var_het}.
Under the assumption of mean homogeneity, the estimated standard deviation for the random  slope is found to be 0.13 for ID: 298. (Keep in mind that the random effect variance is patient-specific since it is scaled by $\sigma_i^2$.)
Relative to that, the estimated slope of -0.43 is clearly unexpected and inconsistent with overall population, leading to the flagging of the profile as mean heterogeneous with probability 0.633.
Identifying such a patient as having an unexpectedly steep drop in her hormone level may have clinical implications worth further investigated.
Profiles with ID 649 and 776 in the first row of Figure \ref{fig:DHEAS_mean_var_het} are treated as heterogeneous due to having moderately large negative random intercepts,
suggesting a consistently low hormone level that is inconsistent with the patient-to-patient variability in the rest of the cohort.
Profile ID 246 is considered borderline with a posterior probability of $\widetilde{u}_i = 0.456 < 0.5$.
The profiles in the second row with least probable mean heterogeneity appear to have  random intercepts and slope both close to zero.

	
Last two rows in Figure \ref{fig:DHEAS_mean_var_het} shows the variance heterogeneity present in the data.
The profiles in the top row appear to have a substantial fluctuations around their respective random effects $\bZ_i\bsb_i$ indicating high residual variability.
It is worth mentioning here that a large residual variance is not necessary to model an outcome trajectory that substantially deviates from the $\bZ_i\bsb_i$ line.
This scenario may also arise from residuals with small variance but a correlation structure that is not consistent with $\bR_i(\rho)$.  The negative autocorrelation in the latter half of ID 104's trajectory contributes to this individual being identified as variance heterogeneous.
The profiles in the bottom row appear to be relatively much flatter.
The overall variance heterogeneity for the data is approximately 2\% as reported in Table \ref{tab:DHEAS_model_selection}.

We conclude the analysis of DHEAS data discussing the effects of the predictors on the heterogeneity indicators. Table \ref{tab:DHEAS_het_out_effect} suggests that none of the variables have a significant effect on the heterogeneity and outlier indicators.
However, the 95\% CI for BMI is substantially shifted towards  positive values in the outlier model indicating that it might have a potential positive effect on the outlier probability,
suggesting that overweight patients may be more likely to have an occasional uncharacteristic lab value. However, high BMI patients are not more likely to have an extreme trend or consistently higher variability than patients with normal BMI.

\begin{table}[!tb]
\centering
\footnotesize
\begin{tabular}{lccccccc}
\toprule
           &  & \multicolumn{2}{c}{Mean Heterogeneity}   & \multicolumn{2}{c}{Outliers}             & \multicolumn{2}{c}{Variance Heterogeneity} \\
Parameters &  & $\widehat{\bgamma}_u$ & 95\% CI          & $\widehat{\bgamma}_w$ & 95\% CI          & $\widehat{\bgamma}_z$   & 95\% CI          \\ \midrule
Intercept & & -3.616 & (-4.316, -2.860) & -3.291 & (-3.783, -2.836) & -3.182 & (-3.804, -2.561)\\
t & & - & - & -0.055 & (-0.220, 0.092) & - & -\\
t$^2$ & & - & - & - & - & - & -\\
t$^3$ & & - & - & - & - & - & -\\
Age & & 0.077 & (-0.106, 0.281) & 0.008 & (-0.142, 0.162) & 0.038 & (-0.154, 0.212)\\
BMI & & 0.032 & (-0.173, 0.216) & 0.145 & (-0.015, 0.306) & 0.048 & (-0.139, 0.229)\\
\bottomrule
\end{tabular}
\caption{The effects of the predictors in heterogeneity and outlier identification are also reported.
}
\label{tab:DHEAS_het_out_effect}
\end{table}

To further demonstrate our method, we provide an additional data analysis in Appendix \ref{appn-sec:cd4} illistrating HOILD on CD4 data from an HIV clinical trial \citep{henryRandomizedControlledDoubleBlind1998a}.
This is an unbalanced longitudinal data, where patients have variable number of measurements recorded at different time points.
See Appendix \ref{appn-sec:cd4} for the complete analysis.


\section{Discussion} \label{sec:discussion}

In this article, we have proposed a comprehensive framework that identifies three different sources of heterogeneity in longitudinal data, alongside providing efficient estimation of the usual target of inference in such settings, the fixed effects coefficients.
Profiles with extreme mean are accounted for by inflating the individual-specific residual variance through an indicator variable specifying subject-level mean heterogeneity.
We account for the outlier observations within a profile by increasing the observation-specific marginal variances by a scaling factor determined by a contamination-style indicator variable.
To identify profiles with an overall high level of fluctuation, individual specific residual variances are modeled using a mixture distribution characterized by a third indicator variable that determines subject-level variance heterogeneity.

In addition to identifying the sources of heterogeneity through these indicators, our models  also consider how the characteristics of an individual can be useful in predicting the extreme behavior by fitting logistic regression models to the indicators.
This enables the identification of specific patient phenotypes associated with subject-level heterogeneity, facilitating more targeted and patient-focused treatment strategies.
In contrast, investigating for observation-level outliers 
can provide a structured approach for 
data validation and quality control.
We consider two different real datasets, one with a fixed number of measurements per individual and another where the numbers vary between individuals (see Appendix \ref{appn-sec:cd4}).
Analysis using our approach shows a meaningful amount of heterogeneity present in both datasets, and we identify patient characteristics that may be informing the heterogeneity indicators.

A classic diagnostic approach for outlier identification is to assess whether a residual is significantly larger than the residual variance.
Building on this, \cite{mccullochImprovingPredictionsWhen2023, mccullochFlaggingUnusualClusters2024} proposed a robust random intercept model that identifies extreme clusters by comparing cluster-level deviations against the residual variance.
Our proposed framework for longitudinal data is more general as it can simultaneously accommodate and identify subject-level extremes in both location and scale, as well as observation-level outliers, while performing robust estimation of fixed-effects.
Another line of research \citep{welsh13ApproachesRobust1997, pinheiroEfficientAlgorithmsRobust2001} uses a scale mixture of normal distributions (marginally $t$ distribution) to provide robust estimation of fixed-effects by down-weighting extreme data points, but this offers only an indirect strategy for identifying extreme patterns.
In contrast, our proposed model allows for direct identification of outliers through superior tail-customization, enabling us to regulate the ``extremeness'' of the heterogeneous group relative to the homogeneous majority through our choice of the tuning parameter $\eta$.

Although our proposal resembles the formulation of a generic mixture model, the two components of each heterogeneity indicator have their own specific interpretations.
Unlike traditional mixture modeling frameworks \citep{quintanaBayesianNonparametricLongitudinal2016, yuMixtureRegressionLongitudinal2022} that partition individuals into homogeneous latent classes, our approach puts individuals showing average behavior into a homogeneous group, and the rest with extreme longitudinal patterns are placed in a single, dispersed heterogeneous group where residual variances are allowed to be subject-specific.
While \cite{pageBayesianLocalContamination2011b} offers a flexible non-parametric contamination model that treats each timepoint as potentially contaminated, it lacks the ability to identify heterogeneity at the subject-level.
Our model provides a more structured alternative by utilizing individual-specific scales that can be directly linked to patient characteristics, ensuring that the identified ``extremeness'' is both statistically robust and clinically interpretable.


We conclude our discussion by highlighting some potential extensions of our framework that can accommodate more complex data structures. 
A current characteristic of our model is the assumption of symmetry, which can be relaxed using proposals available in the literature.
For instance, \cite{sahuNewClassMultivariate2003} used a skew-$t$ and \cite{lachosLIKELIHOODBASEDINFERENCE2010, lachosRobustLinearMixed2009} used skew-normal and a scale mixture of multivariate skew-normal \citep{brancoGeneralClassMultivariate2001} to accommodate thick-tailed skewed random effects and a scale mixture of normal for the errors.
Integrating our approach into such models would facilitate a more comprehensive identification of heterogeneity in non-symmetric clinical data.
Another potential generalization of our model would be to extend it to the setting of generalized linear mixed models (GLMM) with dispersion parameters, such as the gamma distribution.
In this context, the response $y_{ij}$ follows an exponential family distribution $\text{EF}(\mu_{ij}, \phi_{ij})$, where the mean $\mu_{ij} = g(\zeta_{ij})$ is a function of the linear predictor $\zeta_{ij} = \bX_{ij}'\bbeta + \bZ_{ij} \bsb_{i}$ and the dispersion $\phi_{ij}$ is specific to an individual and a timepoint.
Random effects in $\zeta_{ij}$  will account for mean heterogeneity though extreme values of $\bsb_i$ as in the linear case.
The dispersion parameter $\phi_{ij}$ can be taken as a global value $\phi_{ij}=\phi_0$ for non-outliers, variance homogenous observations ($Z_i=0$,$W_{ij}=0$), and appropriately modified depending on the combination of indicators: 
$\phi_{ij}=\eta_w^2  \phi_0$ for $Z_i=0$, $W_{ij}=1$;
$\phi_{ij}=\phi_i$ for $Z_i=1$, $W_{ij}=0$;
$\phi_{ij}=\eta_w^2 \phi_i$ for $Z_i=1$, $W_{ij}=1$.
As in $\sigma_i^2$, the $\phi_i$ for variance heterogeneous cases will come from some distribution chosen to provide dispersion that is on average larger than $\phi_0.$
Specifying a full implementation of this strategy is left for future work.

\section*{Acknowledgements}
Computing resources used in this project was supported in part by the U.S. National Science Foundation (NSF) under grant CNS1828521 and the University of Louisville’s Research Computing team.
Additional support was provided by the National Institutes of Health (NIH) / National Institute of Dental and Craniofacial Research (NIDCR) under Grant No.\ 1R01DE025835 (PI: Jeffrey M.\ Albert, ``Continuous Time Causal Mediation Models for Social Behavior in Health'').





\bibliographystyle{apalike}
{\bibliography{references.bib}}

\newpage
\appendix

\renewcommand{\theequation}{S.\arabic{equation}}
\setcounter{equation}{0}
\renewcommand{\thefigure}{S.\arabic{figure}}
\setcounter{figure}{0}
\renewcommand{\thetable}{S.\arabic{table}}
\renewcommand*{\theHtable}{\thetable}
\setcounter{table}{0}


\section{Posterior Computational Details}
\label{appn-sec:comp_algo}

The MCMC algorithm for the proposed $\hemhevo$ model iterates through the following steps.
\begin{enumerate}

  \item \textit{Updating} $\bgamma_w$: We update $\bgamma_w$ in two steps: first draw $(\omega_{ij} | \bgamma_{w}) \sim \text{PG}(1, \bsx_{w;ij}' \bgamma_w)$ and then update $\bgamma_w$ by drawing from $\text{MVN}_{d_2}(\bmu_w^*, \mathbf{\Sigma}_w^*)$, where \\
  $ \mathbf{\Sigma}_w^* = \left[ c_w^{-2} \bSigma_{w} + \sum_i\sum_j \omega_{ij} \bsx_{w;ij} \bsx_{w;ij}' \right]^{-1} $ and \\
  $ \bmu_w^* = \mathbf{\Sigma}_w^* \left[c_w^{-2} \bmu_w + \sum_i \left( \sum_j w_{ij} - 1/2 \right) \bsx_{w;ij} \right] $.

  
  \item \textit{Updating} $\bgamma_z$: The two steps for updating $\bgamma_z$ are: first draw $(\omega_i|\bgamma_z) \sim \text{PG}(1, \bsx_{z;i}'\bgamma_z)$, then update $\bgamma_z$ by drawing from $\text{MVN}_{d_1}(\bmu^*_z, \mathbf{\Sigma}_z^*)$, where \\
  $ \mathbf{\Sigma}_z^* = \left[ c_z^{-2} \bSigma_z + \sum_i \omega_{i} \bsx_{z;i}\bsx'_{z;i} \right]^{-1} $ and $\bmu_z^* = \mathbf{\Sigma}_z^* \left[ c_z^{-2} \bmu_z + \sum\limits_i \left( z_{i} - 1/2 \right) \bsx_{z;i} \right] $. 

  \item \textit{Updating} $\bgamma_u$: The two steps for updating $\bgamma_u$ are: first draw $(\omega_i|\bgamma_u) \sim \text{PG}(1, \bsx_{u;i}'\bgamma_u)$, then update $\bgamma_u$ by drawing from $\text{MVN}_{d_1}(\bmu_u^*, \mathbf{\Sigma}_u^*)$, where \\
  $ \mathbf{\Sigma}_u^* = \left[ c_u^{-2} \bSigma_u + \sum_i \omega_{i} \bsx_{u;i}\bsx'_{u;i} \right]^{-1} $ and $\bmu_u^* = \mathbf{\Sigma}_u^* \left[ c_u^{-2} \bmu_u + \sum\limits_i \left( u_{i} - 1/2 \right) \bsx_{u;i} \right] $. 
  
  \item \textit{Updating $(U_i, Z_i, \bsW_i)$}: 
  For $i = 1,\ldots, n$ and $j = 1, \ldots, n_i$, sample $(U_i, Z_i, W_{ij} | \bsW_{i(-j)}) \sim \text{Multinomial}(1;\boldsymbol{p}^*_{i})$, where $\boldsymbol{p}^*_{i}$ is a 8-dimensional vector consisting of posterior probabilities corresponding to the eight possible values of $(U_i, Z_i, W_{ij})$. The posterior probability for $(U_i, Z_i, W_{ij}) = (u_i, z_i, w_{ij})$ is denoted by $p^*_{(u_i, z_i, w_{ij})}$, where
  \[
  \begin{array}{rcl}
  p^*_{(u_i, z_i, w_{ij})} & \propto & \left(\pi_{w;ij}^{w_{ij}} (1-\pi_{w;ij})^{1 - w_{ij}} \right) \left(\pi_{u;i}^{u_i} (1-\pi_{u;i})^{1-u_i} \right) \times \\
  & & \begin{cases}
      (1-\pi_{z;i}) f_{0}(\boldsymbol{y}_i | \bbeta, \sigma_0^2, \rho, u_i, \bsw_i), & \text{if } z_i = 0\\
      \pi_{z;i} f_{1}(\boldsymbol{y}_i | \bbeta, \sigma_0^2, \rho, \alpha^2, u_i, \bsw_i), & \text{if } z_i = 1
  \end{cases},
  \end{array}
  \]
  
  where $f_{z0}(\cdot)$ is the density corresponding to the homogeneous group and $f_{z1}(\cdot)$ represents the density corresponding to the heterogeneous group marginalized over $\sigma_i^2$. The two densities are presented in equation (\ref{eq:f0_f1_mixture}) discussed in Section \ref{sec:model_indicators} .


  
  \item \textit{Updating $\sigma_0^2, \sigma_1^2$}: To update $\sigma_0^2$ we need a Metropolis-Hastings (MH) step. We note that the full conditional distribution is proportional to $\pi(\sigma_0^2) \prod_{\{i:z_i = 0\}} f_{0}(\boldsymbol{y}_i) \prod_{\{i:z_i = 1\}} f_{1}(\boldsymbol{y}_i) $, where $f_1(\cdot)$ and $f_2(\cdot)$ are given in equation (\ref{eq:f0_f1_mixture}). The proposal distribution we use for $\sigma_0^2$ is given by, $({\sigma^*_0}^{2} | \sigma_0^2) \sim \text{LN}(\sigma_0^2, \sigma_{\text{MH}}^2) $. For $\sigma_1^2$, the full conditional distribution is proportional to $\pi(\sigma_1^2|\sigma_0^2, \alpha^2) \prod_{i:z_i=1} f_1(y)$ and we use the proposal distribution $({\sigma^*_1}^{2} | \sigma_1^2) \sim \text{LN}(\sigma_1^2, \sigma_{\text{MH}}^2) $.

   \item \textit{Updating $\alpha^2$}: We use a Metropolis-Hastings (MH) step to update $\alpha$. The full conditional distribution f or $\alpha$ is proportional to $\pi(\alpha) \prod_{\{i:z_i = 1\}} f_{1}(\boldsymbol{y}_i) $, where $f_0(\cdot)$ and $f_1(\cdot)$ are defined in equation (\ref{eq:f0_f1_mixture}). The proposal distribution we use for $\alpha^2$ is given by $({\alpha^{2*}} | \alpha^2) \sim \text{LN}(\alpha^2, \sigma_{\text{MH}}^2)$.
  
  \item \textit{Updating $\rho$}: To update $\rho$ we use a Metropolis-Hastings (MH) step. The full conditional distribution for $\rho$ is proportional to $\pi(\rho) \prod_{\{i:z_i = 0\}} f_{0}(\boldsymbol{y}_i) \prod_{\{i:z_i = 1\}} f_{1}(\boldsymbol{y}_i) $, where $f_0(\cdot)$ and $f_1(\cdot)$ are defined in equation (\ref{eq:f0_f1_mixture}). The proposal distribution we use for $\rho$ is given by, $(\rho' | \rho) \sim \text{Unif}(\max\{\rho-\sigma_\rho, 0\}, \min\{\rho+\sigma_\rho, 1\}) $.
  
  \item \textit{Updating $\sigma_{i}^2$}: We update $\sigma_i^2$ conditional on the indicator variable $Z_i = z_i \in \{0,1 \}$. 
  For $i=1,\ldots,n$, we set $\sigma_i^2 = \sigma_0^2$ if $z_i=0$, otherwise $\sigma_i^2$ is drawn from $\text{GIG}(x;m,a,b)$ given by
  \[
  \frac{(a/b)^{m/2}}{2K_m(\sqrt{ab})}\, x^{m-1} \exp \left\{-\frac{1}{2} (ax + b/x) \right\},
  \]
  where $m = 1/\alpha^{2} - n_i/2$, $a = S^2(\bsy_i|\bsw_i)$, and $b = 2/(\alpha^2 \sigma_1^2)$.

  \item \textit{Updating $\bsb_i$}: For $i=1, \ldots, n$, the full conditional distribution for $\bsr_i$ is given by $\MVN(\bmu_{r_i}, \bSigma_{r_i})$, where 
  \[
  \begin{array}{rcl}
  \bSigma_{r_i} & = & \sigma_i^2 \left[ \{(\eta_u^2)^{u_i} \bLambda \}^{-1} + \bZ_i \bOmega^{-1}(\bsw_i) \bZ_i' \right]^{-1} \\
  \bmu_{r_i}^* & = & \bSigma_{r_i} \bZ_i' \left\{\sigma_i^{2} \bOmega(\bsw_i) \right\}^{-1} (\bsy_i - \bX_i \bbeta).
  \end{array}
  \]
  
  \item \textit{Updating $\bLambda$}: The full conditional distribution for $\bLambda$ conditional on $\bsb_i, i=1, \ldots, n$, is given by $\IW \left( \nu + n, \bPsi + \sum_{i=1}^n \sigma_i^{-2} (\eta_u^{-2})^{u_i} \bsb_i \bsb_i' \right)$, where $q$ is the number of random effects.

  \item \textit{Updating $\bbeta$}: The full conditional distribution of $\bbeta$ is $\text{MVN}_p \left( \bmu_{\bbeta}^*, \bSigma_{\bbeta}^* \right)$, where
  \[
  \begin{array}{rcl}
  \bSigma_{\bbeta}^* & = & \left[c_{\bbeta}^{-2} \mathbf{I}_{p} + \sum_{i=1}^n \mathbf{X}_i' (\sigma_i^2 \widetilde{\bOmega}(u_i, \bsw_i))^{-1} \mathbf{X}_i\right]^{-1}, \\
  \mu_{\bbeta}^* & = & \bSigma_{\bbeta}^* \left[ c_{\bbeta}^{-2} \bmu_{\bbeta} + \sum_{i=1}^n \mathbf{X}_i' (\sigma_i^2 \widetilde{\bOmega}(u_i, \bsw_i))^{-1} \boldsymbol{Y}_i \right].
  \end{array}
  \]

\end{enumerate}

We make some comments about adapting the sampler for the special case models.
For model $\homhov$, we keep $u_i$, $w_{ij}$ and $z_i$ fixed at 0 for all $i,j$, and also skip sampling $\gamma_u$, $\gamma_w$ and $\gamma_z$ at all the iterations.
While we sample $\gamma_u$ and $u_i$ for every $i$ under $\hemhov$, $w_{ij}$ and $z_i$ are kept fixed at zero across all iterations, and $\gamma_w$ and $\gamma_z$ are not updated.
Similarly for fitting $\homhovo$, $\gamma_w$ and $w_{ij}$ for every $i$ and $j$ are sampled while $u_i$ and $z_i$ for every $i$ remain fixed at zero and $\gamma_u$ and $\gamma_z$ are not updated.
For model $\hemhovo$, we sample $\gamma_u$, $\gamma_w$ along with $u_i$ and $w_{ij}$ for every $i,j$, and only fix $z_i$ at zero and do not sample $\gamma_z$.

To improve mixing and convergence when using versions of the model with multiple heterogeneity indicators, we introduce these terms incrementally in the model during the burn-in period.
In particular, the sampler begins by fitting the baseline $\homhov$ model, such that all $U_i=W_{ij}=Z_i=0$.  After a short number of iterations to achieve stability, the model shifts gradually through increasingly complex models ($\hemhov$, $\hemhovo$) and indicator variables are turned on, pausing for a couple hundred iterations at each model, until finally sampling from the full model.  This strategy helps ensure that the homogeneous components of the mixtures are correctly centered at the dominant group.

\section{Estimation and Sensitivity to Tuning Parameters}\label{appn-sec:eta}


We recall here that the tuning parameters $\eta_u$, $\eta_w$, and $\eta_z$ regulate the degree of extremeness of the heterogeneous component with respect to the homogeneous component.
More specifically, $\eta_u, \eta_w, \eta_z > 1$, and the larger the $\eta$s are
the more extreme observations or profiles have to be in order to be identified as heterogeneous. 
Necessarily, for a fixed data set, increasing $\eta$ will yield fewer outliers and heterogeneous individuals.  
For this reason, we primarily recommend choosing $\eta$ based what level of difference represents a meaningful discrepancy from the population average in the context of the specific application.  
However, even with a subjectively chosen value of $\eta$, it is good practice to consider the sensitivity to this choice.

As discussed in Section \ref{sec:sim_data_generation}, we generate 100 datasets from each of the models $\hemhov$, $\homhovo$, and $\hemhevo$, under the true values of $\eta_u = \eta_w = \eta_z = \eta = 3$. 
Each dataset from these generative models is fit using all proposed models for each choice of $\eta$ within $\{2,3,4,5\}$.
In the following sections, we will discuss the sensitivity to $\eta$ choice in terms of model selection, estimation of the fixed effects, and heterogeneity identification.

\subsection{Model selection} \label{appn-sec:eta_model_selection}


\begin{table}[!tb]
    \centering
    \footnotesize
    \begin{tabular}[t]{lccccc}
        \toprule
        & M-WAIC & C-WAIC & $\widehat{u}$ & $\widehat{w}$ & $\widehat{z}$ \\
        \midrule
        \multicolumn{6}{l}{{Data,Model: $\hemhov$}}\\
        \hspace{1em}$\eta$: 2 & 0.00 & 0.00 & 0.078 & 0 & 0\\
        \hspace{1em}$\eta$: 3 & 0.96 & 0.52 & 0.077 & 0 & 0\\
        \hspace{1em}$\eta$: 4 & 0.04 & 0.48 & 0.069 & 0 & 0\\
        \hspace{1em}$\eta$: 5 & 0.00 & 0.00 & 0.063 & 0 & 0\\


        
        \midrule
        \multicolumn{6}{l}{{Data,Model: $\hemhevo$}}\\
        \hspace{1em}$\eta$: 2 & 0.23 & 0.00 & 0.063 & 0.016 & 0.076\\
        \hspace{1em}$\eta$: 3 & 0.77 & 0.76 & 0.065 & 0.019 & 0.056\\
        \hspace{1em}$\eta$: 4 & 0.00 & 0.24 & 0.056 & 0.017 & 0.057\\
        \hspace{1em}$\eta$: 5 & 0.00 & 0.00 & 0.045 & 0.016 & 0.060\\

        \midrule
        \multicolumn{6}{l}{{Data,Model: $\homhovo$}}\\
        \hspace{1em}$\eta$: 2 & 0.85 & 0.00 & 0 & 0.017 & 0\\
        \hspace{1em}$\eta$: 3 & 0.15 & 0.24 & 0 & 0.017 & 0\\
        \hspace{1em}$\eta$: 4 & 0.00 & 0.75 & 0 & 0.015 & 0\\
        \hspace{1em}$\eta$: 5 & 0.00 & 0.01 & 0 & 0.014 & 0\\
        \bottomrule
    \end{tabular}
    \caption{Selection of $\eta_u = \eta_w = \eta_z = \eta$ based on M-WAIC and C-WAIC.}
    \label{tab:sim_eta_selection}
\end{table}
\FloatBarrier



Table \ref{tab:sim_eta_selection} evaluates $\eta$ selection performance using M-WAIC and C-WAIC.  As shown in the main manuscript, both M-WAIC and C-WAIC can effectively distinguish between HOILD model variations, so here, we only consider the selection of $\eta$ within the true data generating model. 
When the model is fit with $\eta =$ 2, 3, 4, or 5, the rate of M-WAIC selecting the correct value of $\eta$ is 96\% for the data model $\hemhov$ and 77\% for  $\hemhevo$.
The equivalent rates of making correct choices using C-WAIC are slightly worse at  52\% and 76\%, respectively.
However, selection of $\eta$ is not perfect for the outlier-only datasets generated from $\homhovo$.
While M-WAIC underestimates $\eta$, favoring $\eta=2$ in 85\% of the simulations and selecting the truth with 15\% accuracy, C-WAIC prefers higher value ($\eta = 4$) and achieve a 24\% correct identification rate.
As expected, we observe that as $\eta$ increases, 
$\widehat{u}$, $\widehat{w}$ and $\widehat{z}$, the proportions of subjects/observations classified as heterogeneous decrease.
Since $\eta$ provides a metric for extremeness of the heterogeneous group, a higher $\eta$ will only include more extreme profiles, reducing the estimated rate of heterogeneity in the data.

This suggests that WAIC metrics perform reasonably well, with M-WAIC performing slightly better, in identifying the true $\eta$ in our simulation setting.  However, we continue to recommend that the final selection of $\eta$ for a dataset be guided primarily by subject matter context and informed by the model selection criteria.

\subsection{Fixed effects estimation} \label{appn-sec:eta_fixed_effects}

\begin{figure}[!tb]
\centering
\includegraphics[width=\linewidth]{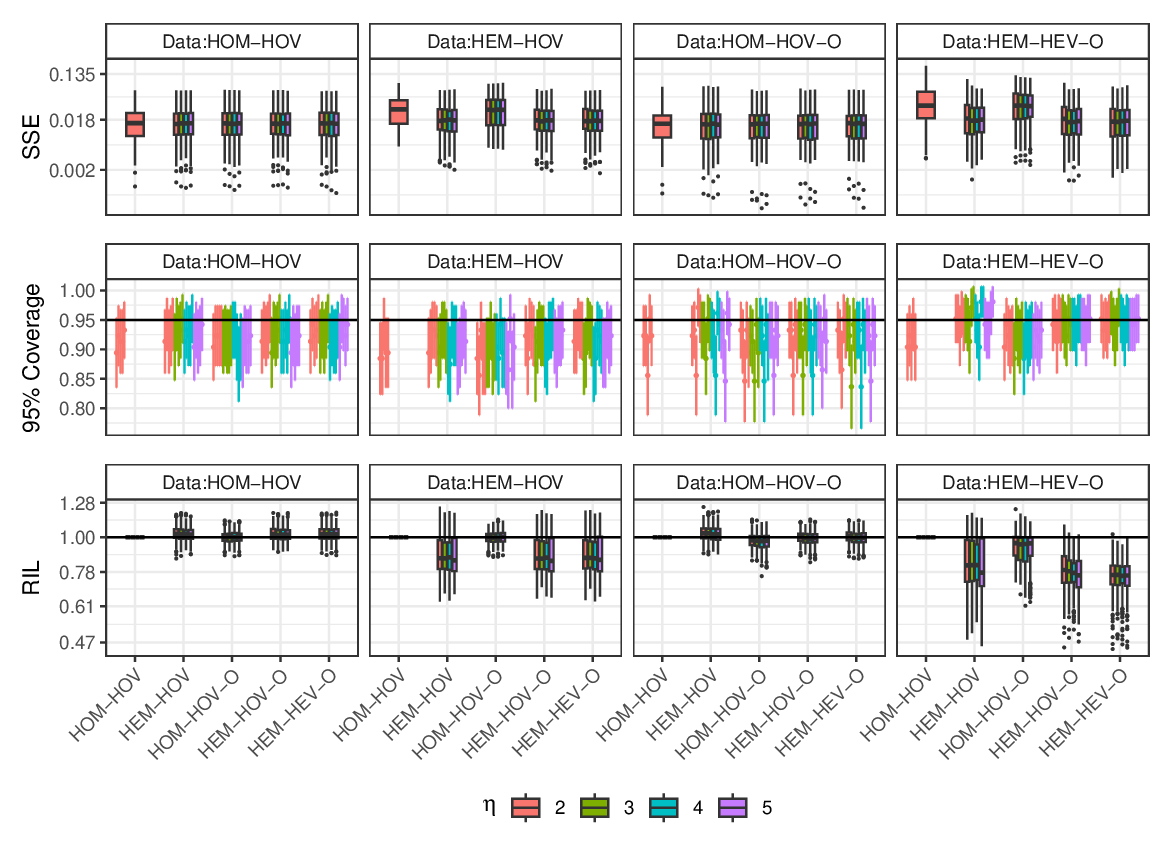}
 \caption{Fixed effects estimation performance across different models in three different data settings. Since $\homhov$ does not involve $\eta$, we have one plot for each metric.}
\label{fig:sim_fixed_eff_eta}
\end{figure}

We now discuss the effect of $\eta$ on $\bbeta$ estimation presented in Figure \ref{fig:sim_fixed_eff_eta}.
We do not observe any meaningful differences between the SSEs and the 95\% coverage rates for different choices of $\eta$ across all data settings, suggesting that inference on the fixed effects are generally robust to the choice of $\eta$ within the suggested range of 2--5.
We also do not observe any meaningful difference between the relative interval lengths (RILs) as $\eta$ varies.

\subsection{Heterogeneity identification} \label{appn-sec:eta_heterogeneity}

\begin{figure}[!tb]
\centering
\includegraphics[width=0.8\linewidth]{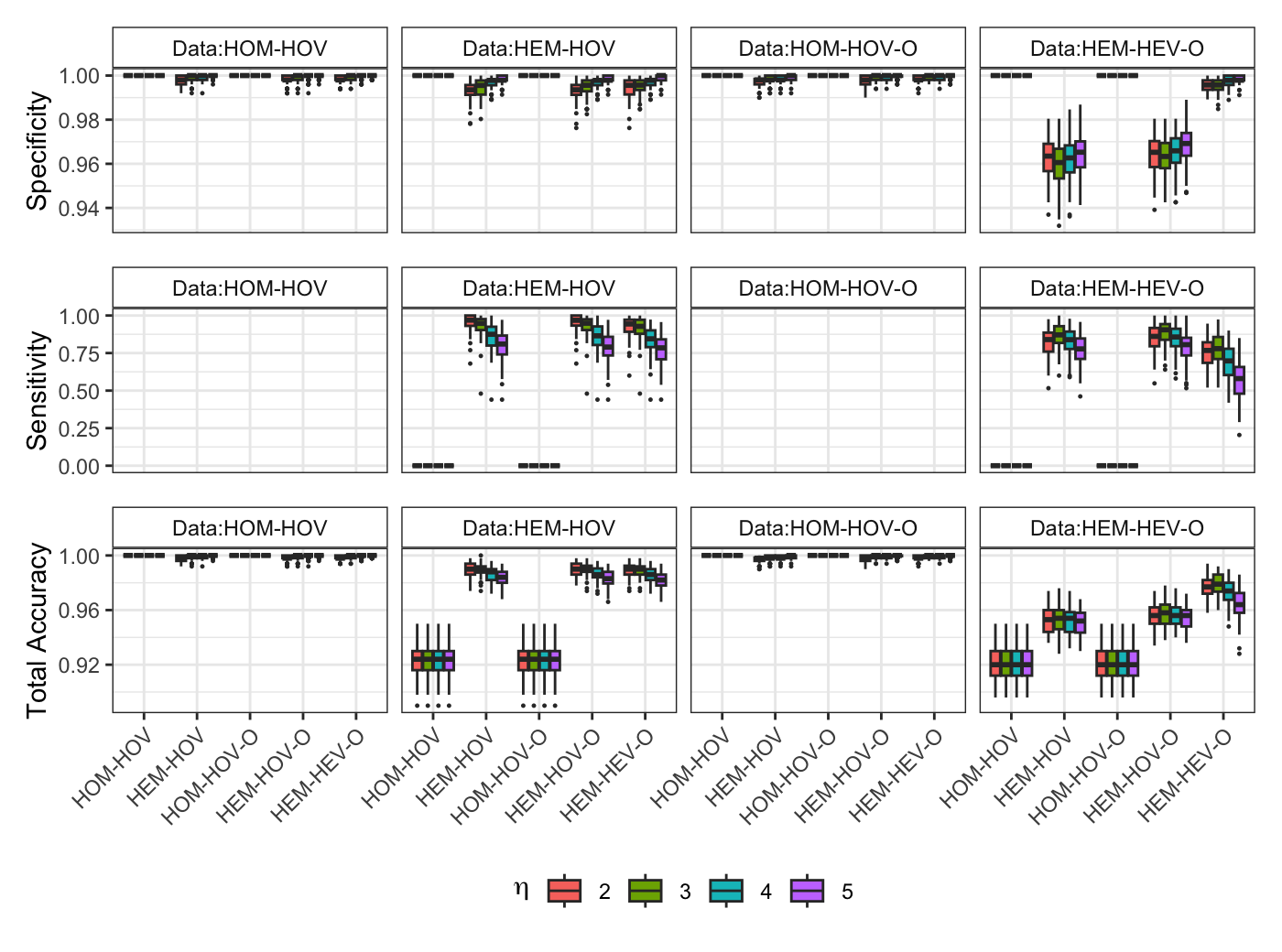}
\caption{The specificity, sensitivity and total accuracy of the subject-level mean heterogeneity indicator $U$ across different models under different data settings.}
\label{fig:sim_u_eta}
\end{figure}

\begin{figure}[!tb]
\centering
\includegraphics[width=0.8\linewidth]{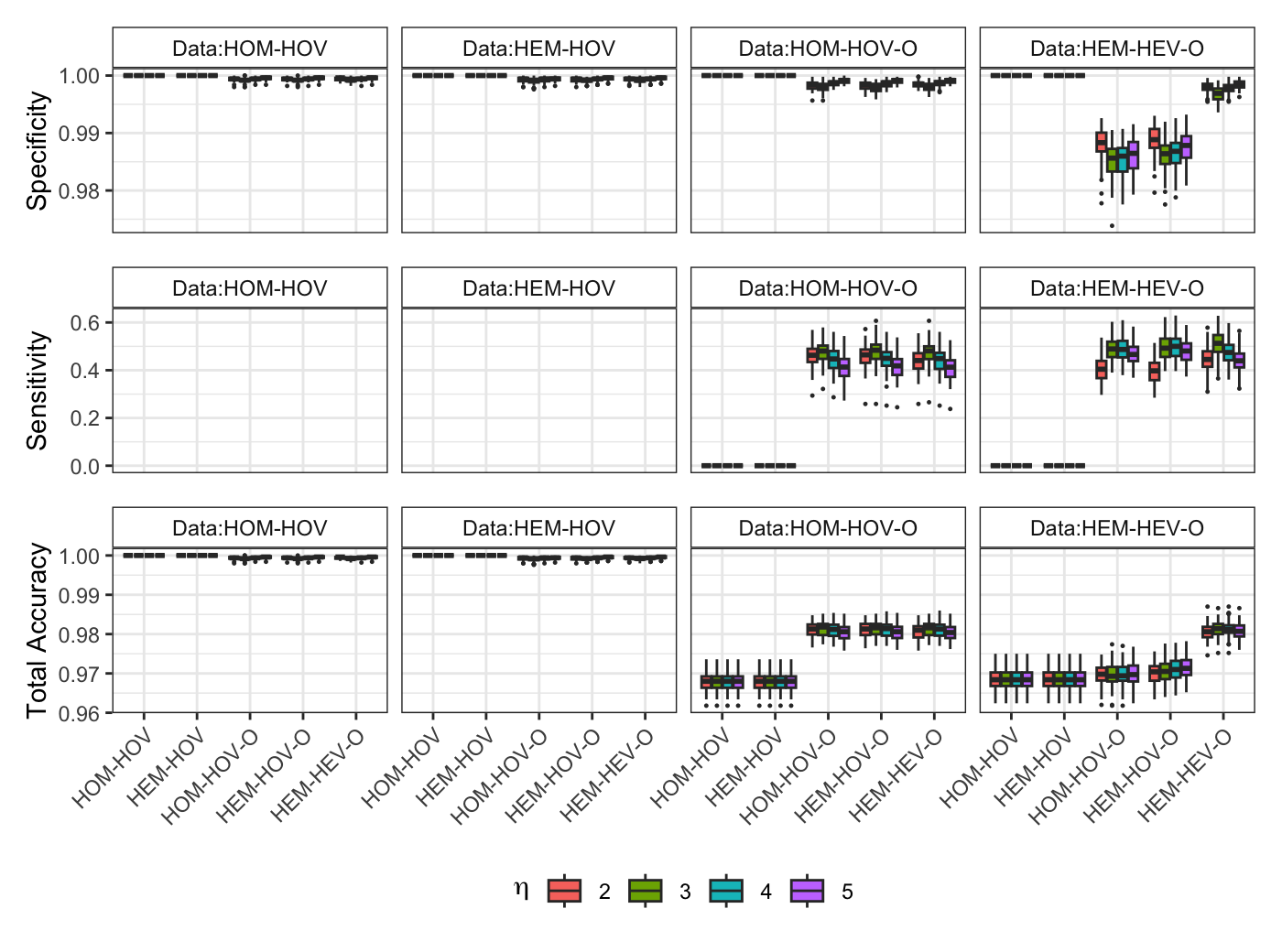}
\caption{The specificity, sensitivity and total accuracy of the observation-level outlier indicator $W$ across different models under different data settings.}
\label{fig:sim_w_eta}
\end{figure}

\begin{figure}[!tb]
\centering
\includegraphics[width=0.8\linewidth]{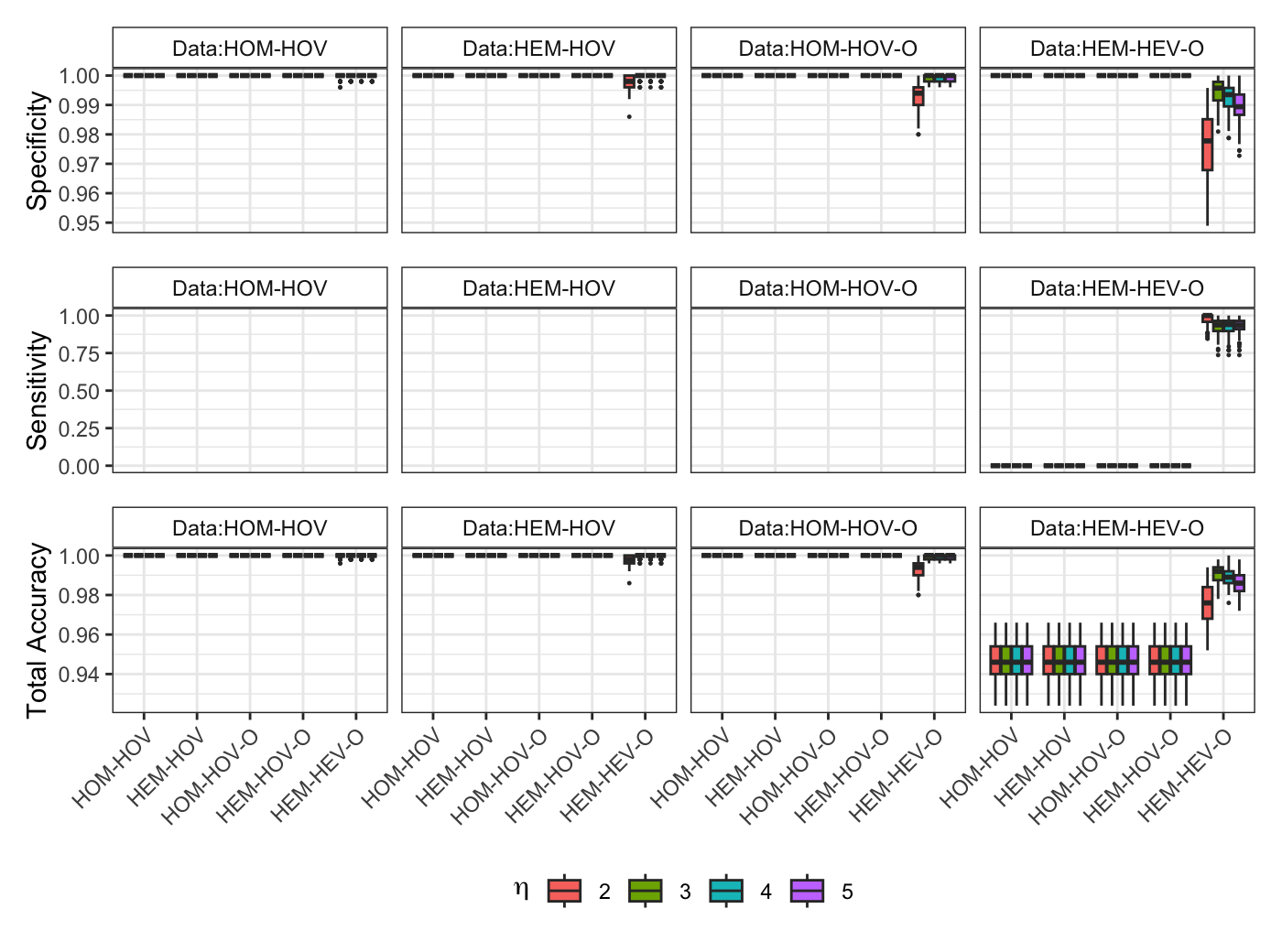}
\caption{The specificity, sensitivity and total accuracy of the subject-level variance heterogeneity indicator $Z$ across different models under different data settings.}
\label{fig:sim_z_eta}
\end{figure}

The influence of $\eta$ on the specificity and sensitivity of heterogeneity identification is presented in Figures \ref{fig:sim_u_eta}, \ref{fig:sim_w_eta}, and \ref{fig:sim_z_eta}.
We observe that, generally, as $\eta$ grows, the specificity increases and the sensitivity decreases.
This is the expected behavior as $\eta$ smaller than the true value will classify some homogeneous profiles as heterogeneous leading to a smaller specificity, while a larger $\eta$ will treat some of the heterogeneous profiles as homogeneous, reducing the identification sensitivity.
An exception to this pattern occurs for $\eta = 2$, particularly under the $\hemhevo$ setting, where high specificity and low sensitivity are observed relative to $\eta > 2$.
This can be attributed to the lack of separation between the homogeneous and heterogeneous groups at small values of the scaling factor $\eta$.
The structural similarity of the two groups at $\eta = 2$ leads to misclassifications of the heterogeneous profiles.
Throughout the three indicator variables, the total accuracy is generally highest at $\eta=3$, consistent with the true data generating parameter.


\FloatBarrier
\newpage
\section{Parameter Estimation under $t$-distributed Data Generation}
\label{appn-sec:fixed_effects_t}

\begin{figure}[!tb]
\centering
\includegraphics[width=0.9\linewidth]{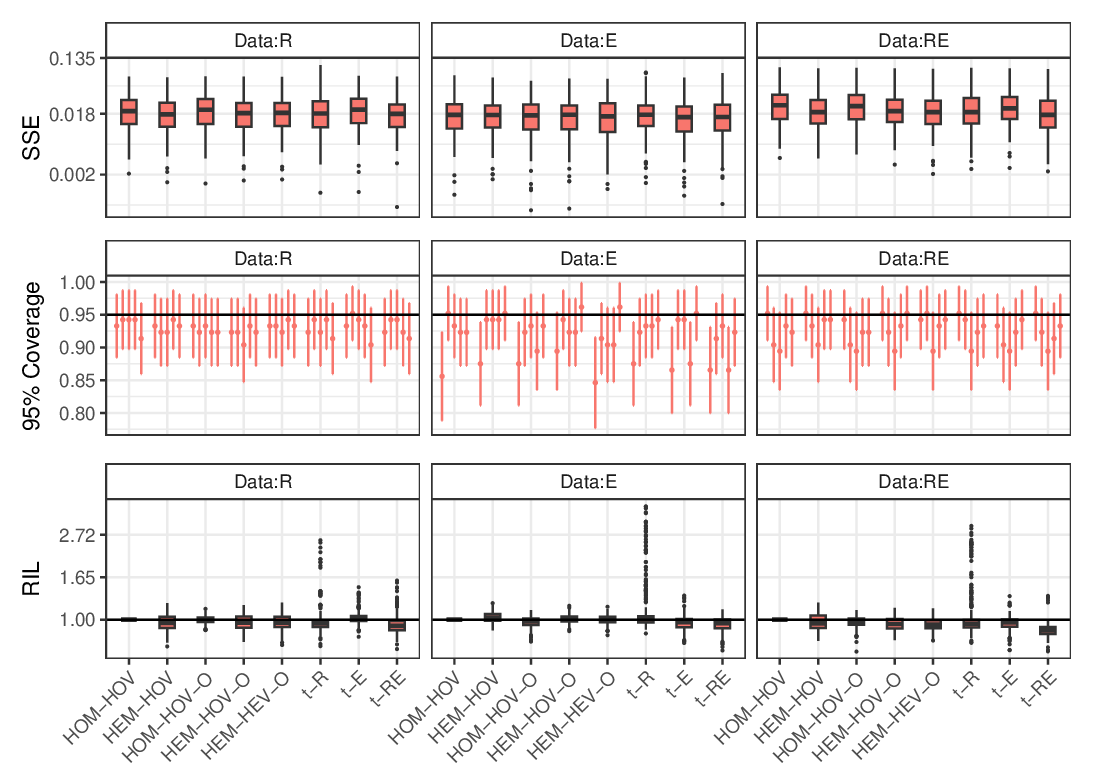}
 \caption{Fixed effects estimation performance across different models in three different data settings.}
\label{fig:sim_fixed_eff_estimation_t}
\end{figure}

Longitudinal models with a thick-tailed distribution are often used to account for extreme observations in the data.
We have compared our proposed framework against models involving multivariate $t$-distributions ($\tR$, $\tE$ and $\tRE$) in Section \ref{sec:sim_fixed_eff}, when the data were generated under our HOILD framework.
In contrast, here the comparison is made using datasets generated under the thick-tailed (non-contamination) models. 
Specifically, we generate 100 datasets from each of the models $\tR$, $\tE$, and $\tRE$ and compare SSE, 95\% coverage, and RIL for each data setting in Figure \ref{fig:sim_fixed_eff_estimation_t}.
Recall that $\tR$ allows for subject-level mean heterogeneity in the data, while $\tE$ accounts for subject-level variance heterogeneity and $\tRE$ explains both.

All models  perform comparably in terms of SSE across all datasets.
Although the models involving $t$-distribution perform well as the true data-generating process, the proposed models that account define heterogeneity through binary indicators also deliver equally good SSE.
In particular, the SSE for $\hemhov$ is similar to that for $\tR$, when the data generative model is also $\tR$.
More complicated models $\hemhovo$, $\hemhevo$ and $\tRE$, which explain mean heterogeneity, also have a similar SSE.
When the data generative model is $\tRE$, we obtain the best SSE with $\hemhevo$ and $\tRE$.
All models across all datasets provide very similar 95\% coverage rate for fixed effects.
The last row of Figure \ref{fig:sim_fixed_eff_estimation_t} shows that the credible intervals for the proposed models are as good as the $t$-distribution models.
Model $\tRE$ provides the narrowest credible intervals across all data scenarios with some extreme RILs occassionally.
In contrast, our proposed models achieve slightly inferior RILs but behave more consistently.
In summary, our proposed framework, while misspecified in these settings, still performs reasonably well compared to true generative models, and better than the standard mixed effect model $\homhov$.
This demonstrates the effectiveness of our approach in a variety of heterogeneous data scenarios.

\section{CD4 Data Example} 
\label{appn-sec:cd4}

The second data set we analyze originally comes  from \cite{henryRandomizedControlledDoubleBlind1998a}
and has been frequently analyzed in other longitudinal applications, including comprehensive case studies by \cite{fitzmauriceAppliedLongitudinalAnalysis2011} and methodological works like \cite{fengStatisticalInferenceHeterogeneous2022, houHeterogeneousQuantileRegression2024}. The data was collected from a randomized double-blind study of AIDS patients with advanced immune suppression.
The study assigned 1309 patients randomly to one of four daily treatment regimens. 
The measurements of CD4 counts were collected at baseline, with follow-up measurements at approximately 8-week intervals. 
One of the main features of this data is that it is unbalanced, i.e.,  the recorded measurements are misaligned and subject to missingness due to skipped visits and dropout. 
The number of measurements per individual varies between 1 and 9. 
Our analysis includes the 779 individuals who had  CD4 counts recorded at baseline and during at least three follow-up visits.

\begin{figure}[!tb]
    \centering
    \includegraphics[width=0.8\textwidth]{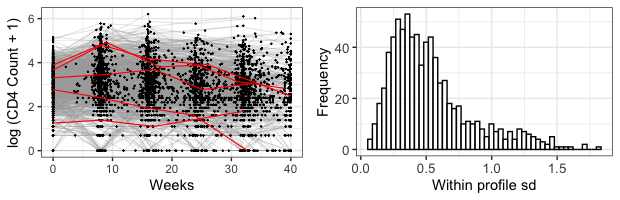}
    \caption{The figure shows the distribution of log-transformed CD4 counts data.}
    \label{fig:CD4}
\end{figure}

We transformed the CD4 counts as log(CD4 counts + 1) for the outcome variable. The following covariates were chosen as potential predictors: 
(1) a quadratic function of time $t$, measuring the number of weeks from the first visit with $t=0$ representing the baseline, 
(2) age, 
(3) gender, and 
(4) the four treatment regimes --- (a) zidovudine alternating monthly with 400mg considered as control, (b) zidovudine plus 2.25mg of zalcitabine, (c) zidovudine plus 400mg of didanosine, and (d) zidovudine plus 400mg of didanosine plus 400mg of nevirapine.  The treatments were encoded using three binary dummy variables. 
We again chose $\eta_u = \eta_w = \eta_z \in \{3, 4, 5\}$. 
Figure \ref{fig:CD4} shows an overview of the CD4 data that clearly indicates heterogeneity in the data.

\begin{table}[!h]
\centering
\footnotesize
\begin{tabular}[t]{lccccccc}
\toprule
Model($\eta_u, \eta_w, \eta_z$) & M-WAIC & C-WAIC & $\widehat{u}$ & $\widehat{w}$ & $\widehat{z}$ & $\widehat{\sigma_0^2}$ & 95\% CI \\
\midrule
$\homhov$ & 17813.5 & 17813.5 & 0 & 0 & 0 & 0.335 & (0.315, 0.353)\\
\addlinespace
$\hemhov$ (3,-,-) & 17848.7 & 17838.8 & 0.003 & 0 & 0 & 0.316 & (0.299, 0.333)\\
$\hemhov$ (4,-,-) & 17855.5 & 17837.9 & 0.003 & 0 & 0 & 0.317 & (0.3, 0.334)\\
$\hemhov$ (5,-,-) & 17856.8 & 17834.4 & 0.003 & 0 & 0 & 0.318 & (0.301, 0.336)\\
\addlinespace
$\homhovo$ (-,3,-) & 17395.6 & 15940.4 & 0 & 0.029 & 0 & 0.154 & (0.131, 0.174)\\
$\homhovo$ (-,4,-) & 17695.4 & 15722.7 & 0 & 0.029 & 0 & 0.142 & (0.117, 0.169)\\
$\homhovo$ (-,5,-) & 18186.9 & 15773.5 & 0 & 0.024 & 0 & 0.147 & (0.118, 0.171)\\
\addlinespace
$\hemhovo$ (3,3,-) & 17341.9 & 15893.1 & 0.006 & 0.032 & 0 & 0.137 & (0.118, 0.160)\\
$\hemhovo$ (4,4,-) & 17560.1 & 15622.3 & 0.005 & 0.034 & 0 & 0.121 & (0.099, 0.141)\\
$\hemhovo$ (5,5,-) & 18057.4 & 15693.8 & 0.001 & 0.027 & 0 & 0.130 & (0.106, 0.148)\\
\addlinespace
$\hemhevo$ (3,3,3) & 17227.0 & 15933.7 & 0.000 & 0.021 & 0.062 & 0.121 & (0.097, 0.141)\\
$\hemhevo$ (4,4,4) & 17516.5 & 15750.3 & 0.000 & 0.028 & 0.009 & 0.115 & (0.095, 0.138)\\
$\hemhevo$ (5,5,5) & 18009.4 & 15847.8 & 0.000 & 0.023 & 0.013 & 0.123 & (0.101, 0.144)\\
\bottomrule
\end{tabular}
\caption{Estimation performance of different models for CD4 data with different $\eta$'s in terms of WAIC, rates of three heterogeneity types, and homogeneous group residual variance $\sigma_0^2$.}
\label{tab:CD4_model_selection}
\end{table}

Table \ref{tab:CD4_model_selection} contains model selection results across the HOILD variations.  We note here that the inclusion of the mean heterogeneity alone $\hemhov$ leads to a slightly worse model fit that the standard mixed effects model $\homhov$.
In contrast, accommodating observation-level outliers helps to achieve a better fit for model $\hemhovo$ indicated by smaller M-WAIC.
The estimated $\sigma_0^2$ is also much smaller than before, and the overall proportion of outliers is approximately 3\% out of the total 3834 observations.
Considering variance heterogeneity in the full model $\hemhevo$, the M-WAIC improves slightly, and we chose the $\hemhevo$ model with $\eta = 3$ to be the optimal model based on M-WAIC.

\begin{table}[!tb]
\footnotesize
\centering
\begin{tabular}{lccccccc}
\toprule
           &  & \multicolumn{2}{c}{$\homhov$}           &  & \multicolumn{3}{c}{$\hemhevo$}                \\
                         &  & \multicolumn{2}{c}{(mixed effects model as reference)}           &  & \multicolumn{3}{c}{(full heterogeneous selected model)}                \\
Parameters &  & $\widehat{\bbeta}$ & 95\% CI          &  & $\widehat{\bbeta}$ & 95\% CI          & RIL  \\ \midrule
Intercept & & 2.871 & (2.674, 3.108) & & 2.836 & (2.632, 3.064) & 1.00\\
Week & & 0.063 & (-0.001, 0.127) & & 0.034 & (-0.022, 0.085) & 0.84\\
Week$^2$ & & -0.222 & (-0.293, -0.164) & & -0.192 & (-0.247, -0.146) & 0.78\\
Age & & 0.090 & (0.033, 0.147) & & 0.096 & (0.036, 0.161) & 1.10\\
Zidovudine+Zalcitabine & & 0.130 & (-0.038, 0.305) & & 0.159 & (-0.016, 0.319) & 0.98\\
\addlinespace
Zidovudine+Didanosine & & 0.152 & (-0.022, 0.320) & & 0.211 & (0.042, 0.379) & 0.99\\
Zidovudine+Didanosine+ & & 0.025 & (-0.138, 0.199) & & 0.074 & (-0.093, 0.229) & 0.95\\
\hspace{6em}Nevirapine & & & & & & & \\
Gender & & -0.054 & (-0.266, 0.135) & & 0.023 & (-0.171, 0.227) & 0.99\\
\bottomrule
\end{tabular}
\caption{Fixed effects of the predictors for CD4 data. RIL=relative interval length represents the ratio of the interval length of $\hemhevo$ to $\homhov$.
}
\label{tab:CD4_preds_effect}
\end{table}


Table \ref{tab:CD4_preds_effect} reports the fixed effect estimates for the proposed model $\hemhevo$ with $\eta_u = \eta_w = \eta_z = 3$ and compares them against the simplest model $\homhov$ with no heterogeneity.
While most of the effects have similar estimates, there is a meaningful change in the inference regarding the treatment effects under the model that accounts for heterogeniety.
Although the credible interval for Zidovudine + Didanosine has a trend towards a positive association with CD4 count in the homogeneous model, this effect becomes statistically significant when we allow heterogeneity in the model (in the sense of the credible interval excluding zero).
Note that we gain efficiency with $\hemhevo$ for all predictor effects (except Age) irrespective of their significance.
The gain is most apparent for the time predictors, where we observe 16\% reduction in RIL for the week variable and 22\% reduction for week$^2$.
Although neither of the models suggests any significant effect for the first and third treatments, both treatments have marginally positive associations with CD4 count with greater evidence of a treatment effect in the $\hemhevo$ analysis.

\begin{figure}[!tb]
  \centering
  \includegraphics[width=0.8\textwidth]{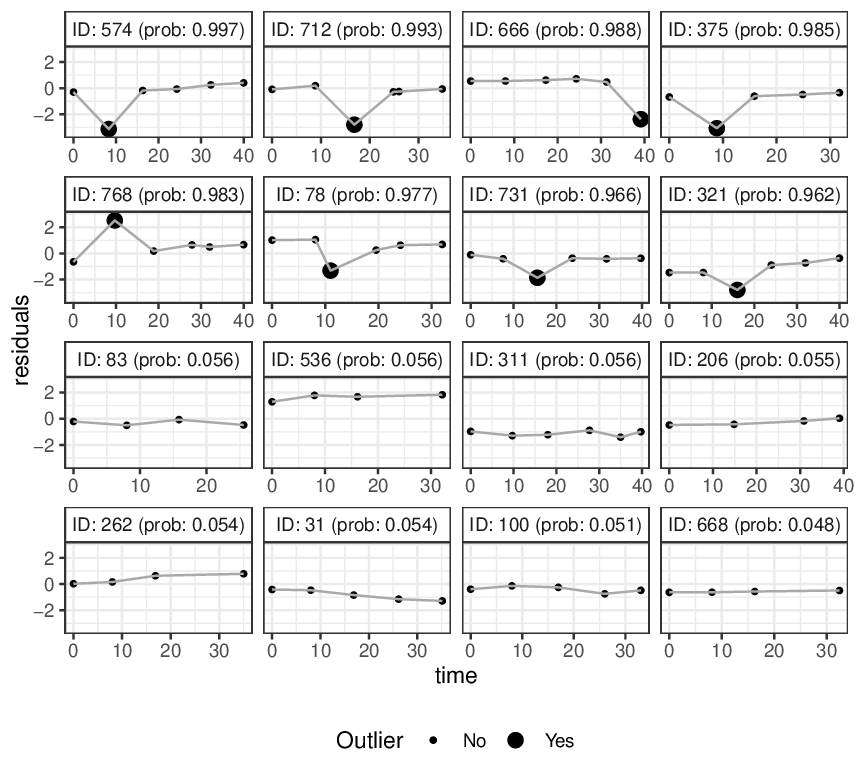}
  \caption{Fixed effects residual profiles ($\hat{\beps}_i = \bsy_i - \bX_i\hat{\bbeta}$) with larger dots representing outliers, and the headers report the corresponding probabilities $\tilde{w}_{ij}$. The profiles are in decreasing order of the probablities, the top-left subfigure displaying the trajectory of subject ID 574, whose $\log(\text{CD4}+1)$ count at $t=2$ has the highest probability of being an outlier among all the outcome values across all the subjects.}
  \label{fig:CD4_outlier}
\end{figure}

\begin{figure}[!tb]
  \centering
  \includegraphics[width=0.8\textwidth]{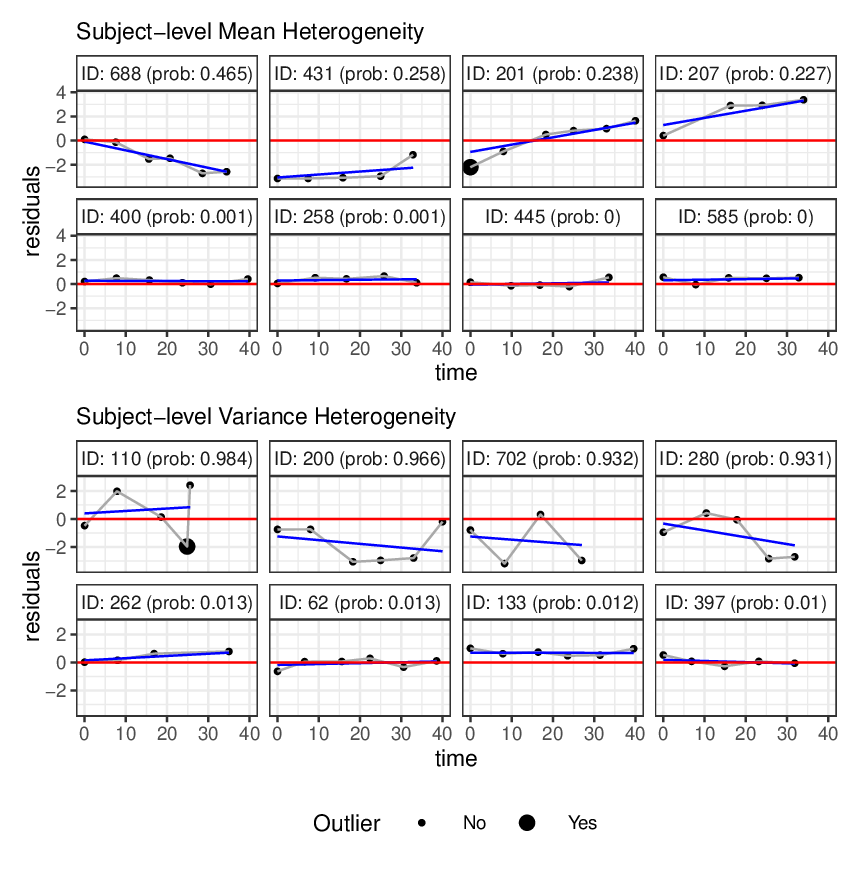}
  \caption{Fixed effects residual profiles ($\hat{\beps}_i = \bsy_i - \bX_i\hat{\bbeta}$) with subject-level mean and variance heterogeneity. 
		First two rows show profiles with most and least probable mean heterogeneous profiles respectively and the bottom two rows show profiles with most and least variance heterogeneity.
		The blue line shows the random effects $\bZ_i\bsb_i$. 
		The probabilities for the relevant heterogeneity are reported in the header of each subfigure.}
  \label{fig:CD4_mean_var_het}
\end{figure}

We illustrate the outlier identification by our proposed model in Figure \ref{fig:CD4_outlier}.
The top two rows show visible inconsistency of the outliers compared to the rest of their respective profiles while the bottom two rows have no prominent outliers.
Figure \ref{fig:CD4_mean_var_het} shows profiles with subject-level mean and variance heterogeneity.
Only the first individual represents any potential mean heterogeneity with a borderline posterior probability of 47\%, driven primarily by a steep and decreasing slope.
The third-row profiles show considerable fluctuations around the blue line representing the mean random effects, with their respective probabilities greater than 0.9.
The profiles in the bottom row have smaller residual variance and are treated as variance homogeneous with probability exceeding 98\%. 

Finally, we report the predictor effects relevant for the identification of heterogeneity and outliers in Table \ref{tab:CD4_het_out_effect}.
We observe that none of treatments show much association with subjects displaying either  mean and variance heterogeneity.
However, we do see that male gender tends to have higher rates of outliers, while time is borderline negatively associated with outliers.
Patients are less likely to have an outlier measurement the longer they have been in the trial, and the ability of our model to identify the time-varying presences of outliers is likely associated with the increased efficiency of the fixed effects for time.

\begin{table}[!tb]
\centering
\footnotesize
\begin{tabular}{lcccccc}
\toprule
           & \multicolumn{2}{c}{Mean Heterogeneity}   & \multicolumn{2}{c}{Outliers}             & \multicolumn{2}{c}{Variance Heterogeneity} \\
Parameters & $\widehat{\bgamma}_u$ & 95\% CI          & $\widehat{\bgamma}_w$ & 95\% CI          & $\widehat{\bgamma}_z$   & 95\% CI          \\ \midrule
Intercept & -3.771 & (-4.388, -3.021) & -2.386 & (-3.012, -1.810) & -2.351 & (-3.110, -1.711)\\
Week & - & - & -0.134 & (-0.285, 0.020) & - & -\\
Week$^2$ & - & - & - & - & - & -\\
Age & -0.005 & (-0.202, 0.168) & 0.002 & (-0.149, 0.157) & 0.020 & (-0.151, 0.210)\\
Zidovudine+Zalcitabine & -0.012 & (-0.205, 0.184) & 0.061 & (-0.133, 0.242) & 0.031 & (-0.160, 0.220)\\
Zidovudine+Didanosine & -0.014 & (-0.201, 0.183) & 0.070 & (-0.119, 0.248) & 0.034 & (-0.150, 0.225)\\
Zidovudine+Didanosine+ & -0.022 & (-0.225, 0.169) & 0.082 & (-0.112, 0.261) & 0.017 & (-0.180, 0.194)\\
\hspace{6em}Nevirapine & & & & & & \\
Gender & -0.055 & (-0.245, 0.139) & 0.235 & (0.035, 0.466) & 0.054 & (-0.164, 0.232)\\
\bottomrule
\end{tabular}
\caption{Predictors in heterogeneity and outlier identification for CD4 data.
}
\label{tab:CD4_het_out_effect}
\end{table}

\end{document}